\documentclass[prb,twocolumn,superscriptaddress,showpacs]{revtex4-1}

\usepackage{amsmath}    %
\usepackage{graphicx}  %
\usepackage{bm}        %
\usepackage{amssymb}   %

\usepackage{color}

\newcommand{\bra}{\langle}
\newcommand{\ket}{\rangle}

\begin{document}

\title{Self-consistent DFT+U method for real-space time-dependent density functional theory calculations}%

\author{Nicolas Tancogne-Dejean}
  \email{nicolas.tancogne-dejean@mpsd.mpg.de}
  \affiliation{Max Planck Institute for the Structure and Dynamics of Matter, 
                 Luruper Chaussee 149, D-22761 Hamburg, Germany}
 \affiliation{European Theoretical Spectroscopy Facility (ETSF)}
 
 \author{Micael J.~T.~Oliveira}
  \email{micael.oliveira@mpsd.mpg.de}
  \affiliation{Max Planck Institute for the Structure and Dynamics of Matter, 
                 Luruper Chaussee 149, D-22761 Hamburg, Germany}
  \affiliation{European Theoretical Spectroscopy Facility (ETSF)}

 \author{Angel Rubio}
  \email{angel.rubio@mpsd.mpg.de}
  \affiliation{Max Planck Institute for the Structure and Dynamics of Matter, 
                 Luruper Chaussee 149, D-22761 Hamburg, Germany}
 \affiliation{European Theoretical Spectroscopy Facility (ETSF)}
 \affiliation{Nano-Bio Spectroscopy Group, Universidad del Pa\'is Vasco, 20018 San Sebasti\'an, Spain}

\begin{abstract}
We implemented various DFT+U schemes, including the ACBN0 self-consistent density-functional version of the DFT+U method [Phys. Rev. X 5, 011006 (2015)] within the massively parallel real-space time-dependent density functional theory (TDDFT) code Octopus. 
We further extended the method to the case of the calculation of response functions with real-time TDDFT+U and to the description of non-collinear spin systems.
The implementation is tested by investigating the ground-state and optical properties of various transition metal oxides, bulk topological insulators, and molecules.
Our results are found to be in good agreement with previously published results for both the electronic band structure and structural properties. The  self consistent calculated values of U and J are also in good agreement with the values commonly used in the literature.
We found that the time-dependent extension of the self-consistent DFT+U method yields improved optical properties when compared to the empirical TDDFT+U scheme.
This work thus opens a different theoretical framework to address the non equilibrium properties of correlated systems.
\end{abstract}

\maketitle
\section{Introduction}

It was recognize very early that the simplest local and semilocal approximations of density-functional theory (DFT) fail to describe the electronic and structural properties of correlated materials, like transition metal oxides.\cite{anisimov1997first,himmetoglu2014hubbard}
Along the years, the DFT+U method originally proposed by V. Anisimov, A. Lichestein, and coworkers\cite{anisimov_band_1991,anisimov_density-functional_1993,liechtenstein_density-functional_1995,anisimov1997first} has become a well established and successful way to improve the treatment of correlated solids upon DFT.
The physical motivation of the DFT+U method, most often referred to as LDA+U, stems from the over-delocalization of the electrons when treated within the local density approximation (LDA)~\cite{PhysRevB.45.13244} or generalized gradient approximations (GGA).~\cite{PhysRevLett.77.3865} In particular, the localization of 3d and 4f electrons turns out to be extremely important for ground-state and excited-state properties of transition metals.
In order to correct this over-delocalization, it was proposed to include an energy penalty, called $U$ in reference to the notation used in Hubbard Hamiltonians, for these orbitals.~\cite{anisimov_band_1991,anisimov_density-functional_1993,liechtenstein_density-functional_1995,anisimov1997first,himmetoglu2014hubbard}
The success of the DFT+U method mainly originates from the simplicity of the method, its relative low computational cost, and the fact that it can predict the proper magnetic ground state of Mott insulators.\cite{anisimov1997first}
However, the method has some intrinsic deficiencies, such as yielding infinite life-times for quasi-particles, or opening the band-gap by making a long-range order. Therefore, methods going beyond the DFT+U method, that is, going beyond the mean-field level, such as the DMFT~\cite{PhysRevLett.62.324,PhysRevB.45.6479} or more recently DFT+DMFT~\cite{RevModPhys.78.865}, have become the state-of-the-art methods to treat strongly correlated materials.
These methods have been developed in depth over the last years, at the price of an higher computational cost than DFT-type methods. As such, the DFT+U approach still remains very attractive when it comes to the calculation of larger systems.

In its original formulation, the DFT+U method is not \textit{ab initio}, as no formal recipe was proposed to obtain the Hubbard $U$ and Hund $J$ parameters.
Recently, various schemes have been proposed to obtain the value of the Hubbard $U$ \textit{ab initio}, instead of adjusting the values of $U$ and $J$ to spectroscopic data. 
Among them one finds linear response calculations in super-cells,~\cite{PhysRevB.71.035105} constrained random-phase approximation,~\cite{miyake_screened_2008,aichhorn_dynamical_2009,miyake_ab_2009,aichhorn_theoretical_2010} 
or direct approximations of the (screened) Hartree-Fock energy.~\cite{Agapito_PRX,PhysRevB.76.155123}
In this context, the work of Agapito \textit{et al.}~\cite{Agapito_PRX} is highly relevant as they introduced an efficient pseudo-hybrid functional, the ACBN0 functional, to obtain the $U$ and $J$ values \textit{ab initio} and self-consistently.
This is achieved, in principle, with only a small increase of the computational cost when compared to standard LDA or GGA calculations.
It was shown that the self-consistency in the calculation of the Hubbard $U$ can be crucial in the case of transition-metal complexes.~\cite{PhysRevLett.97.103001}
Furthermore, having a method that allows to dynamically compute the value of the Hubbard $U$ is of fundamental relevance to describe correlated systems out of equilibrium.

This paper is organized as follow.
First we briefly present the usual DFT+U method and the ACBN0 functional along with details of our implementation in the real-space code Octopus~\cite{MARQUES200360, castro_octopus:_2006, andrade_real-space_2015}.
Then, we present numerical results obtained with our implementation for the ground-state of a bulk topological insulator in Sec.~\ref{sec:ACBN0}, the energy and forces of the isolated O$_2$ molecule in  Sec.~\ref{sec:forces}, and for optical properties of two transition metal oxides in Sec.~\ref{sec:results}. 
Finally, we draw our conclusions in Sec.~\ref{sec:conclusions}.

\section{DFT+U in real-space}
\label{sec:DFT_U}

In essence, the DFT+U method replaces the DFT total energy functional $E_{\mathrm{DFT}}[n]$ by the DFT+U total energy functional of the form
\begin{equation}
 E_{\mathrm{DFT+U}}[n,\{n_{mm'}^{I,\sigma}\}] = E_{\mathrm{DFT}}[n] + E_{ee}[\{n_{mm'}^{I,\sigma}\}] - E_{dc}[\{n_{mm'}^{I,\sigma}\}]\,,
 \label{eq:E_DFT_U}
\end{equation}
where $E_{ee}$ is the electron-electron interaction energy, and $E_{dc}$ accounts for the double counting of the electron-electron interaction already present in $E_\mathrm{DFT}$.
Although an exact form of the double-counting term was recently proposed in the context of DFT+DMFT,~\cite{PhysRevLett.115.196403}  
this double-counting term is not known in the general case and several approximated forms have been proposed along the years. 

The $E_{ee}$ and $E_{dc}$  energies depend on the density matrix of a localized orbitals basis set $\{\phi_{I,m}\}$, which are localized orbitals attached to the atom $I$. In the following we refer to the elements of the density matrix of the localized basis as occupation matrices, and we denote them $\{n_{mm'}^{I,\sigma}\}$.
In the rotational-invariant form of DFT+U proposed by Dudarev \textit{et al.},~\cite{PhysRevB.57.1505} one has
\begin{eqnarray}
 E_{ee}[\{n_{mm'}^{I,\sigma}\}] &=& \frac{U}{2} \sum_{m,m',\sigma}N_{m}^{\sigma}N_{m'}^{-\sigma}+\frac{U-J}{2} \sum_{m\neq m',\sigma}N_{m}^{\sigma}N_{m'}^{\sigma}\,, \nonumber\\
  E_{dc}[\{n_{mm'}^{I,\sigma}\}] &=& \frac{U}{2}N(N-1) - \frac{J}{2}N(\frac{N}{2}-1)\,,\nonumber
\end{eqnarray}
where $N=N^{\uparrow}+N^{\downarrow}$ and $N^{\sigma}=\sum_{m} n_{mm}^{\sigma}$.
Combining these two expressions, we obtain the $E_U$ energy  to be added to the DFT total energy, which only depends on an effective  
Hubbard U parameter $U^{\mathrm{eff}}=U-J$,
\begin{eqnarray}
 E_U[\{n_{mm'}^{I,\sigma}\}] = E_{ee}[\{n_{mm'}^{I,\sigma}\}] - E_{dc}[\{n_{mm'}^{I,\sigma}\}]\nonumber\\ 
  = \sum_{I,n,l} \frac{U^{\mathrm{eff}}_{I,n,l}}{2} \sum_{m,\sigma}\Big(n_{mm}^{I,n,l,\sigma} - \sum_{m'}n_{mm'}^{I,n,l,\sigma}n_{m'm}^{I,n,l,\sigma} \Big)\,,
  \label{eq:E_Dudarev}
\end{eqnarray}
where $I$ is an atom index, $n$, $l$ and $m$ refer to the principal, azimuthal, and angular quantum numbers, respectively, and $\sigma$ is the spin index.
In the case of a periodic system, the occupation matrices $n^{I,n,l,\sigma}_{mm'}$ are given by
\begin{equation}
n^{I,n,l,\sigma}_{mm'} = \sum_{n}\sum_{\mathbf{k}}^{\mathrm{BZ}} w_\mathbf{k}f_{n\mathbf{k}}^\sigma \bra\psi_{n,\mathbf{k}}^{\sigma} | \phi_{I,n,l,m}\ket  \bra \phi_{I,n,l,m'}|\psi_{n,\mathbf{k}}^{\sigma} \ket\,, 
\label{eq:occ_matrices}
\end{equation}
where $w_{\mathbf{k}}$ is the $\mathbf{k}$-point weight and $f_{n\mathbf{k}}^\sigma$ is the occupation of the Bloch state $|\psi_{n,\mathbf{k}}^{\sigma} \ket$. 
Here, $| \phi_{I,n,l,m}\ket$ are the localized orbitals that form the basis used to describe electron localization. 

From this energy, the corresponding set of generalized Kohn-Sham equations is obtained by minimizing Eq.~\eqref{eq:E_DFT_U} with respect to the wavefunctions for fixed occupations, which reads as~\cite{PhysRevB.60.10763,PhysRevB.23.5048,PhysRevB.18.7165}
\begin{equation}
 \frac{\delta E_{\mathrm{DFT+U}}[n,\{n_{mm'}^{I,\sigma}\}]}{\delta (\psi^\sigma_{i})^*} - \frac{\delta \sum_{j} E_{j} f_{j} \bra \psi_{j} |\psi_{j} \ket  }{\delta (\psi^\sigma_{i})^*} = 0\,,
 \label{eq:minimization}
\end{equation}
where $i$ and $j$ refer to both band and $\mathbf{k}$-point indexes in the case of periodic systems.
By comparing the resulting equations with the usual Kohn-Sham equations, one obtains the expression of the (nonlocal) potential that must be added to the DFT Hamiltonian.
This potential reads as
\begin{equation}
V^\sigma_{U}|\psi_{n,\mathbf{k}}^{\sigma} \ket = \sum_{I, n,l} \sum_{m,m'} V_{m,m'}^{I,n,l,\sigma} P^{I,n,l}_{m,m'}|\psi_{n,\mathbf{k}}^{\sigma} \ket\,,
\label{eq:pot_V_U}
\end{equation}
where we defined 
\begin{eqnarray}
V_{m,m'}^{I,n,l,\sigma} &=& U^{\mathrm{eff}}_{I,n,l}( \frac{1}{2}\delta_{mm'} - n^{I,n,l,\sigma}_{mm'} )\,,\nonumber\\
P^{I,n,l}_{mm'} &=& |\phi_{I,n,l,m}\ket\bra\phi_{I,n,l,m'}|\,.\nonumber
\end{eqnarray}

\begin{figure}[t]
  \begin{center}
    \includegraphics[width=0.9\columnwidth]{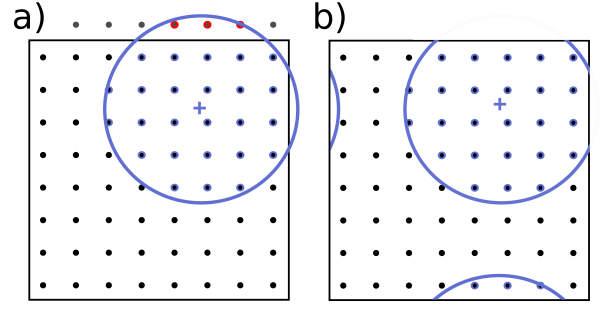}
    \caption{\label{fig:sketch} a) A localized orbital, represented on a sub-grid (violet circle), overlaps with the border of the simulation box. Red points indicate grid points that belong to the sub-grid of the localized orbital, but not to the regular real-space grid. b) The modified sub-grid of the orbital, which is compatible with the periodicity of the solid and with the real-space grid. A phase shift is added to the points which are originally outside of the simulation box. }
  \end{center}
\end{figure}

One important point of the implementation of the DFT+U method lies in the choice and the representation of the localized basis. 
In the Octopus code, which is a norm-conserving pseudopotential-based code, the natural choice is to use the pseudo-atomic wavefunctions.
The pseudo-wavefunctions are usually decomposed into a radial part and an angular part given by the usual spherical harmonics, but 
can easily be expressed on the real-space grid. 
Regarding this, it is worthwhile to note that some care is needed when treating periodic systems.
Indeed, in bulk materials, the orbitals can spread over several unit cells. Therefore, one must at the same time describe an orbital bigger than the size of the simulation box, and preserve periodicity for the main real-space grid. 
We solved this issue by representing the localized orbitals on spherical sub-grids, centered around atoms, in a very similar way to the treatment of the Kleinman-Bylander projectors entering the expression of the non-local representation of the pseudopotentials. 
This is illustrated in Fig.~\ref{fig:sketch}.
Finally, when computing projections of Bloch wavefunctions onto the localized atomic-orbital basis, a phase correction must be added to properly treat the periodicity when  atomic orbitals cross the border of the simulation box.

Ultimately, the radii of the atomic spheres should be big enough to contain the entire localized orbitals. 
This can, however, correspond to a prohibitive number of real-space grid points and in some cases the radii of the atomic spheres must be reduced to obtain a good compromise between accuracy and computational efficiency.
We implemented three methods to choose the radii of these atomic spheres. 
In the first method the atomic spheres are truncated at the radius of the non-local part of the pseudopotential, as proposed in Ref.~\onlinecite{PhysRevB.87.085108}. 
The second method consists in truncating the radii to the smallest dimension of the unit-cell. 
In the last method the radius of the spheres is chosen at the point where the radial component of the pseudo-wavefunctions becomes smaller than some given threshold.

\begin{figure}[t]
  \begin{center}
    \includegraphics[width=0.9\columnwidth]{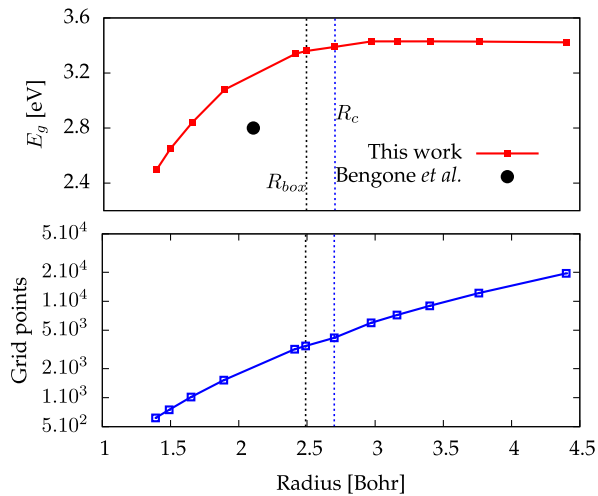}
    \caption{\label{fig:Conv_U5_J095_Radius} Top panel: Convergence of the band-gap of bulk NiO \textit{versus} the truncation radius of the $d$ orbitals of nickel. We used $U_{\mathrm{eff}}=U-J$ with $U=5.0$ eV and $J=0.95$ eV, as chosen in Ref.~\onlinecite{PhysRevB.62.16392}. The back dot corresponds to the value obtained in Ref.~\onlinecite{PhysRevB.62.16392} for an atomic sphere radius of $2.1$ Bohr. 
    Bottom panel: number of grid points covered by the $d$ orbitals of Ni. Two dashed vertical lines indicate respectively the cutoff radius $R_c$ of the non-local part of the pseudopotential used and the smallest dimension of the simulation box $R_{box}$.}
  \end{center}
\end{figure}

In order to illustrate the effect of such truncation, we studied the change of the band-gap of bulk NiO in its anti-ferromagnetic phase, with respect to the radius of the $3d$ orbital of Ni, for a fixed value of $U^{\mathrm{eff}}$.
We used a value of 4.1704~\AA\ for the lattice parameter of NiO, with a real-space spacing of $\Delta r=0.3$ bohr, and a $8\times8\times8$ $\mathbf{k}$-point grid to sample the Brillouin zone. We used the PBE functional~\cite{PhysRevLett.77.3865} for the (semi-)local DFT part.
Our results, presented in Fig.~\ref{fig:Conv_U5_J095_Radius}, show that the band-gap of NiO described using DFT+U strongly depends on the radius of the sphere used for the $3d$ orbitals of the Ni atoms.
Interestingly, restricting the orbitals to a too small sphere results into an underestimation of the band gap which, in turn, might be compensated by an over-estimation of the value of the effective Hubbard $U$. 
As a comparison, we also report from the literature the band gap of NiO computed for the same value of $U^{\mathrm{eff}}$, using the projector augmented wave (PAW) method.~\cite{PhysRevB.62.16392} 
Most of the PAW  and linear muffin-tin orbitals (LMTO) implementations of the DFT+U method~\cite{PhysRevB.62.16392,PhysRevB.60.10763,amadon__2008,liechtenstein_density-functional_1995} are based of the so-called atomic-sphere approximation (ASA).
In a pseudopotential-based code, such as the Octopus code, this approximation is not necessary. 
However, in the spirit of ASA, one could decide to truncate the localized orbitals to the radius of the nonlocal part of the pseudopotential.
Our results reveal that restricting the localized orbitals to the non-local part of the pseudopotential, as proposed in Ref.~\onlinecite{PhysRevB.87.085108}, can yield a good estimate of the band-gap for a moderate number of grid points. However, it is worthwhile to note that the cut-off radius of a pseudopotential strongly depends on the method used to generate it. 
For instance, for nickel one can generate reasonable pseudopotentials whose cut-off radius varies from 1.5 to 2.5 bohr. 
In the following, our results are always converged in total energy with respect to the size of the atomic spheres.

\section{\textit{Ab initio} DFT+U:\\The ACBN0 functional}
\label{sec:ACBN0}

In Ref.~\onlinecite{Agapito_PRX}, an approximation to the electron interaction energy named ACBN0 functional is proposed, allowing for an efficient \textit{ab initio} evaluation of the DFT+U energy, and therefore of the DFT+U Hamiltonian. 
In particular, only the computation of a reduced number of Coulomb integrals is needed to evaluate the effective Hubbard $U$. 
In the following, we always assume orthogonality of the localized basis set attached on each atom. 
This is obviously the case in our implementation, as the norm-conserving atomic pseudo-wavefunctions are orthogonal by construction. Expressions corresponding to the general (non-orthogonal) case are already presented in Ref.~\onlinecite{Agapito_PRX} and are therefore not reported here.
The electron interaction energy is given for the ACBN0 functional by~\cite{Agapito_PRX} 
\begin{eqnarray}
E_{ee} = \frac{1}{2}\sum_{\{m\}}\sum_{\alpha,\beta} \bar{P}_{mm'}^\alpha\bar{P}_{m''m'''}^\beta (mm'|m''m''') \nonumber\\
- \frac{1}{2}\sum_{\{m\}}\sum_\alpha \bar{P}_{mm'}^\alpha\bar{P}_{m''m'''}^\alpha (mm'''|m''m') \,,
\label{eq:HF_ACBN0}
\end{eqnarray}
corrected here for a small typographical mistake (the minus sign between the two terms in Eq.~\eqref{eq:HF_ACBN0}) from the original paper.
In Eq.~\eqref{eq:HF_ACBN0}, the renormalized occupation matrices $\bar{P}^{I,n,l,\sigma}_{mm'}$ and the renormalized occupations $\bar{N}^{I,n,l,\sigma}_{\psi_{n\mathbf{k}}}$ are respectively given by
\begin{equation}
\bar{P}^{I,n,l,\sigma}_{mm'} = \sum_{n\mathbf{k}} w_\mathbf{k}f_{n\mathbf{k}} \bar{N}^{I,n,l,\sigma}_{\psi_{n\mathbf{k}}}  \bra \psi_{n\mathbf{k}}^\sigma |  \phi_{I,n,l,m}  \ket  \bra \phi_{I,n,l,m'} | \psi_{n\mathbf{k}}^\sigma \ket \,,
\label{eq:notation_Pbar}
\end{equation}
  \begin{equation}
  \bar{N}^{I,n,l,\sigma}_{\psi_{n\mathbf{k}}} = \sum_{\{I\}} \sum_{m} \bra \psi_{n\mathbf{k}}^\sigma |  \phi_{I,n,l,m}  \ket  \bra \phi_{I,n,l,m} | \psi_{n\mathbf{k}}^\sigma \ket \,,
  \label{eq:notation_Nbar}
 \end{equation}
where the sums in Eq.~\eqref{eq:notation_Nbar} run over all orbitals of the system owning the quantum numbers $n$ and $l$, and being attached to atoms of the same type as the atom $I$, as this quantity is similar to the Mulliken charge of atom $I$.~\cite{Agapito_PRX} 
From Eq.~\eqref{eq:HF_ACBN0}, the effective Hubbard $U$ is given by $U^{\mathrm{eff}}=\bar{U}-\bar{J}$, where~\cite{Agapito_PRX}
\begin{equation}
\bar{U} = \frac{\sum_{\{m\}}\sum_{\alpha\beta}\bar{P}^\alpha_{mm'}\bar{P}^\beta_{m'm''}(mm'|m''m''')}{\sum_{m\neq m'}\sum_{\alpha} N_{m}^{\alpha}N_{m'}^{\alpha} + \sum_{\{m\}}\sum_{\alpha} N_{m}^{\alpha}N_{m'}^{-\alpha}}\,,
\label{eq:ACBN0_U}
\end{equation}

\begin{equation}
\bar{J} = \frac{\sum_{\{m\}}\sum_{\alpha}\bar{P}^\alpha_{mm'}\bar{P}^\alpha_{m'm''}(mm'''|m''m')}{\sum_{m\neq m'}\sum_{\alpha} N_{m}^{\alpha}N_{m'}^{\alpha} }\,,
\label{eq:ACBN0_J}
\end{equation}
with $N_{m}^{\alpha} = n_{mm}^{I,n,l,\alpha}$. 
Here we omitted the atom and quantum numbers subscripts for conciseness. Altogether, these equations define the ACBN0 functional.
It is important to note that $0\leq \bar{N}^{I,n,l,\sigma}_{\psi_{n\mathbf{k}}} \leq 1$. In the limit of $\bar{N}^{I,n,l,\sigma}_{\psi_{n\mathbf{k}}}=0$, we recover the DFT energy, whereas for  $\bar{N}^{I,n,l,\sigma}_{\psi_{n\mathbf{k}}}=1$ we have the Hartree-Fock energy. However, the energy given by the ACBN0 function lies in between, which is a usually a prerequisite for obtaining reliable electronic gaps.~\cite{PhysRevLett.100.146401}

The potential corresponding to the ACBN0 functional should in principle be obtained by solving Eq.~\eqref{eq:minimization} using Eqs.~\eqref{eq:E_Dudarev},~\eqref{eq:ACBN0_U}, and~\eqref{eq:ACBN0_J}.
This leads to an orbital dependent potential, which must be obtained 
using the optimized effective potential (OEP) method.~\cite{PhysRev.90.317,PhysRevA.14.36,PhysRevLett.90.043004}
Here, however, we decided to implement the functional as defined in the original ACBN0 paper, where the usual DFT+U potential of Eq.~\eqref{eq:pot_V_U} is used.

In Tab.~\ref{tab:comp_ACBN0}, we report the values of the effective $U$ obtained with our implementation, compared with the results of the original ACBN0 paper.~\cite{Agapito_PRX} We also report in Tab.~\ref{tab:comp_ACBN0_gap} the band gaps of transition metal oxides obtained by our implementation, compared to the results obtained in the original ACBN0 paper~\cite{Agapito_PRX} and to experimental results. 
We employed a real-space grid spacing of 0.25 bohr, and a $8\times8\times8$ $\mathbf{k}$-point grid to sample the Brillouin zone. We considered TiO$_2$ in its rutile phase with experimental lattice constants ($a=b=4.594$\AA, $c=2.959$\AA~and $\mu=0.305$ ~\cite{doi:10.1063/1.1676569}),  ZnO in its wurtzite phase with lattice constants ($a=b=3.1995$\AA, $c=5.1330$\AA~and $\mu=0.3816$ ~\cite{doi:10.1063/1.1676569,Agapito_PRX}). As for MnO, we considered it in its ideal rock-salt structure, as already done above for NiO, with type-II antiferromagnetic spin ordering and with a lattice constant of $a=4.4315$\AA. In all the cases, we considered localization on both the $3d$ orbitals of the transition metal atoms and the $2p$ orbitals of the oxygen atoms.

\begin{figure}[t]
  \begin{center}
    \includegraphics[width=0.9\columnwidth]{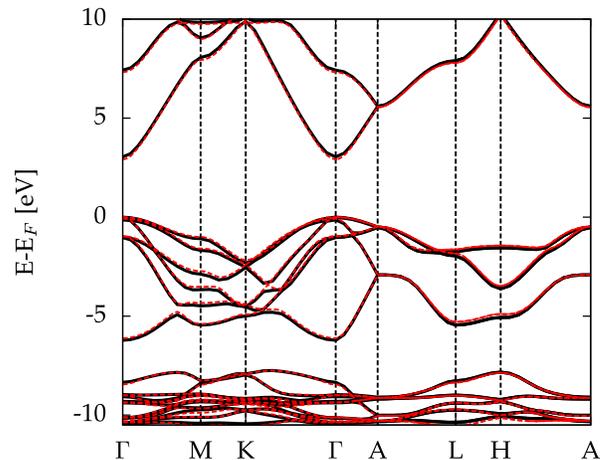}
    \caption{\label{fig:BS_ZnO} Band structure of wurtzite zinc oxide, calculated from our implementation (solid black lines) compared to the result of the original ACBN0 paper (dashed red lines).~\cite{Agapito_PRX}}
  \end{center}
\end{figure}

\begin{figure}[t]
  \begin{center}
    \includegraphics[width=0.9\columnwidth]{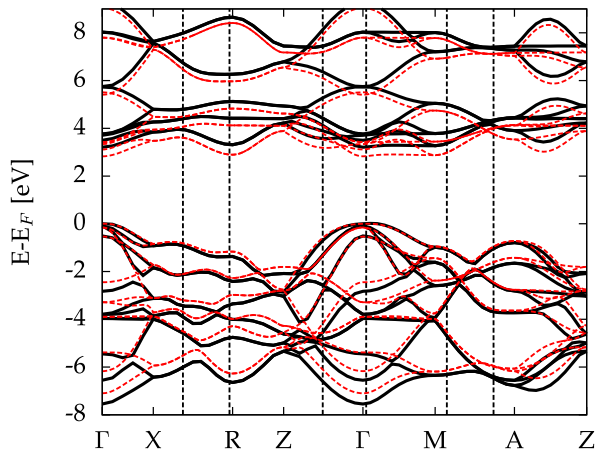}
    \caption{\label{fig:BS_TiO2}  Same as for Fig.~\ref{fig:BS_ZnO}, but for TiO$_2$ in its rutile phase.}
  \end{center}
\end{figure}

Overall, our results are found to be in good agreement with the ones reported in Ref.~\onlinecite{Agapito_PRX}.
However, it is worth to note that one of the difficulties when implementing the ACBN0 functional lies in the calculation of the Coulomb integrals. 
In Ref.~\onlinecite{Agapito_PRX}, the authors make used of a 3G Gaussian basis set to help the calculation of the Coulomb integrals, as it is not possible to compute them directly in reciprocal space. 
A possible origin for discrepancies can therefore be attributed to a different treatment of the localized orbitals, which are not approximated here by a sum of three Gaussians, as we express them directly on the real-space grid.
We also compare the band structures of ZnO and TiO$_2$ calculated from our implementation to the results of Ref.~\onlinecite{Agapito_PRX}, in Figs.~\ref{fig:BS_ZnO} and~\ref{fig:BS_TiO2}, showing a good agreement.
\begin{table}[h]
\begin{ruledtabular}
\begin{tabular}{l|c|c|c|c}
Material & \multicolumn{2}{c|}{TM-3d } & \multicolumn{2}{c}{ Oxygen 2p} \\
         & This work & Ref.~\onlinecite{Agapito_PRX} & This work & Ref.~\onlinecite{Agapito_PRX}  \\
         \hline
MnO & 4.68 & 4.67 & 5.18 & 2.68\\
NiO & 6.93 & 7.63 & 2.87 & 3.0\\
TiO$_2$ & 0.96 & 0.15 & 10.18 & 7.34\\
ZnO & 13.3 & 12.8 & 5.95 & 5.29\\
\end{tabular}
\end{ruledtabular}
\caption{\label{tab:comp_ACBN0}Values of $U_{\mathrm{eff}}$, in eV, for various transition metal oxides, obtained from our implementation, compared to the result of the original ACBN0 paper.~\cite{Agapito_PRX} }
\end{table}

\begin{table}[h]
\begin{ruledtabular}
\begin{tabular}{l|c|c|c}
Material & This work & Ref.~\onlinecite{Agapito_PRX} & Exp.\\
         \hline
MnO  &  2.65 & 2.83 & 4.1; 3.9$\pm$0.4; 3.8-4.2; 3.6-3.8\\
NiO  &  4.14 & 4.29 & 4.0; 4.3; 3.7; 3.7; 3.87\\
TiO$_2$ & 3.21 & 2.83 & 3.3$\pm$0.5; 3.03\\
ZnO  &  3.04 & 2.91 & 3.3. 3.44; 3.44\\
\end{tabular}
\end{ruledtabular}
\caption{\label{tab:comp_ACBN0_gap}Values of the direct band-gap, in eV, for various transition metal oxides, obtained from our implementation, compared to the result of the original ACBN0 paper and to experimental values (see Ref.~\onlinecite{Agapito_PRX} and references therein).}
\end{table}


\begin{figure}[t]
  \begin{center}
    \includegraphics[width=0.9\columnwidth]{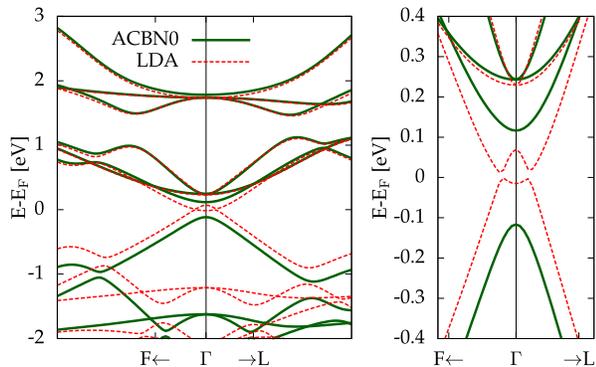}
    \caption{\label{fig:Sb2Te3_noSOC} Left panel: band structure of bulk topological insulation Sb$_2$Te$_3$, calculated using the LDA functional (dashed red lines) and using the ACBN0 functional (green solid lines). Right panel: details of the band-structure around the $\Gamma$ point. }
  \end{center}
\end{figure}

We found that the ACBN0 functional yields improved electronic structures for various types of materials, such as three-dimensional topological insulators.
First we report in Fig.~\ref{fig:Sb2Te3_noSOC} the band structure of Sb$_2$Te$_3$, which is a bulk topological insulator, computed without spin-orbit coupling (SOC).
In this case we employed HGH pseudopotentials~\cite{HGH_pseudos}, a real-space grid spacing of 0.35 bohr, a $8\times8\times8$ $\mathbf{k}$-point grid to sample the Brillouin zone, and used the experimental constant for the lattice parameters and internal constants \cite{wyckoff1963crystal}.
Similarly to previously reported results,~\cite{PhysRevB.88.045206} we find that the LDA band structure exhibits an avoided crossing at the Fermi energy close to the $\Gamma$ point (see right panel of Fig.~\ref{fig:Sb2Te3_noSOC}). 
By including an effective Hubbard $U$ for the $5p$ orbitals of Te in our description, computed from the ACBN0 functional, we open the band-gap of this material. We obtained values of $U_{\mathrm{eff}} = 2.95$ eV and  $U_{\mathrm{eff}} = 2.80$ eV for the Te atom at the Wyckoff position $3a$, and for the two other equivalent Te atoms (Wyckoff position $6c$), respectively. 
The calculated band-gap is $0.23$ eV, in reasonable agreement with the reported full GW band gap of $0.10$ eV.~\cite{PhysRevB.88.045206} It is worth noting here that a one-shot GW calculation leads, for this material, to and unphysical band dispersion close to the Fermi energy, and off-diagonal matrix elements must be included in the self-energy to  correctly dehybridizes the bands (see Ref.~\onlinecite{PhysRevB.88.045206} for a detailed discussion).

In the previous results the SOC was neglected. However, the topological nature of Sb$_2$Te$_3$ and related materials originates from the spin-orbit interaction. 
We have therefore extended the ACBN0 functional to the case of noncollinear spin in order to be able to describe such materials.
The main formulas of DFT+U in the case of noncollinear spin are presented in Appendix~\ref{app_dftu_nc} and the extension of the ACBN0 functional with noncollinear spin is presented in Appendix~\ref{app_acbn0_nc}.
In Fig.~\ref{fig:Sb2Te3_SOC} we compare the band-structure of Sb$_2$Te$_3$ computed with SOC using the LDA and ACBN0 functionals.
Again, we find that our results are in good agreement with the GW results of Ref.~\onlinecite{PhysRevB.88.045206} for the same material, and that the unphysical band dispersion around the $\Gamma$ point for LDA disappears using the ACBN0 functional.
We obtained values of $U_{\mathrm{eff}} = 2.89$ eV and  $U_{\mathrm{eff}} = 2.78$ eV for the Te atom at the Wyckoff position $3a$, and for the two other equivalent Te atoms (Wyckoff position $6c$), respectively. These values are very similar to the values obtained without SOC.
The corresponding band-gap is $0.21$ eV, in excellent agreement with the reported full GW with SOC band gap of $0.21$ eV of Ref.~\onlinecite{PhysRevB.88.045206}. 
\begin{figure}[t]
  \begin{center}
    \includegraphics[width=0.9\columnwidth]{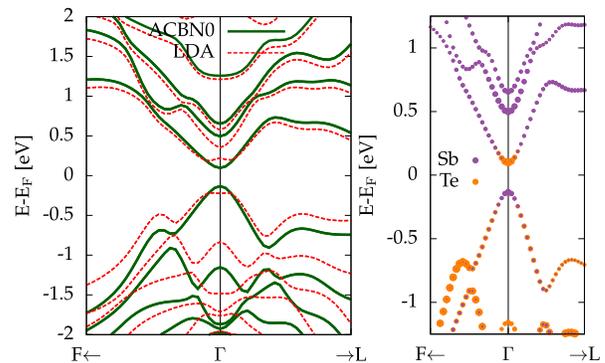}
    \caption{\label{fig:Sb2Te3_SOC} Left panel: same as the left panel of Fig.~\ref{fig:Sb2Te3_noSOC}, but now including full spin-orbit coupling (see text for details of the dealing with spinors in the ACBN0 functional). Right panel: $p$-orbital character of the band structure around the $\Gamma$ point.}
  \end{center}
\end{figure}
As shown in the right panel of Fig.~\ref{fig:Sb2Te3_SOC}, a band inversion around $\Gamma$ point is found with the ACBN0 functional, as expected from the topological nature of Sb$_2$Te$_3$, and in agreement with previous LDA and GW results~\cite{PhysRevB.88.045206}.
Our results indicate that the ACBN0 functional seems to be a good alternative to the computationally expensive full GW calculation for describing bulk topological insulators such as Sb$_2$Te$_3$.

\section{Forces and isolated systems}
\label{sec:forces}

In many relevant physical situations, the knowledge of forces is important. 
Due to the modification of the total energy of the system done in Eq.~\eqref{eq:E_DFT_U}, a contribution to the forces, sometimes referred as Hubbard forces, must be computed.
The contribution to the forces acting on the atom $\alpha$ in the direction $i$ coming from the Hubbard energy of the atom $I$ and the quantum number $n$ and $l$ is defined by
\begin{eqnarray}
F^U_{\alpha,i} = -\frac{\partial E_U}{\partial R_{\alpha,i}} 
= - \sum_{I,m,m',\sigma}  \frac{\partial E_U}{\partial n_{mm'}^{I,n,l,\sigma}}\frac{\partial n_{mm'}^{I,n,l,\sigma}}{\partial R_{\alpha,i}} \nonumber\\
= - \frac{U}{2}\sum_{I,m,m',\sigma} (\delta_{mm'}-2n_{mm'}^{I,n,l,\sigma})\frac{\partial n_{mm'}^{I,n,l,\sigma}}{\partial R_{\alpha,i}} \,.
\end{eqnarray}

From the definition of the occupation matrices for periodic systems, one finds that
\begin{eqnarray}
\frac{\partial n_{mm'}^{I,\sigma}}{\partial R_{\alpha,i}}  = \sum_{\mathbf{k},v}w_{\mathbf{k}}f_{\mathbf{k},v} \Bigg[  \frac{\partial \bra \phi_{m,\mathbf{k}}^{I,n,l}|\psi_{\mathbf{k},v}^\sigma\ket}{\partial R_{\alpha,i}}\bra \psi_{\mathbf{k},v}^\sigma|\phi_{m,\mathbf{k}}^{I,n,l}\ket 
\nonumber\\
+
  \bra \phi_{m,\mathbf{k}}^I|\psi_{\mathbf{k},v}^\sigma\ket \frac{\partial\bra\psi_{\mathbf{k},v}^\sigma| \phi_{m,\mathbf{k}}^I\ket}{\partial R_{\alpha,i}}\Bigg]\,.
\end{eqnarray}

Instead of using directly this expression, we compute the forces from the derivatives of the orbitals in order to reduce the so called egg-box effect.~\cite{andrade_real-space_2015}
After some algebra, one can show that
\begin{eqnarray}
\frac{\partial n_{mm'}^{I,n,l,\sigma}}{\partial R_{\alpha,i}}  = \delta_{\alpha,I} \sum_{\mathbf{k},v}w_{\mathbf{k}}f_{\mathbf{k},v} \Bigg[  \bra \phi_{m',\mathbf{k}}^{I,n,l}|\frac{\partial \psi_{\mathbf{k},v}^\sigma}{\partial r_{i}}\ket\bra \psi_{\mathbf{k},v}^\sigma|\phi_{m,\mathbf{k}}^{I,n,l}\ket 
\nonumber\\
+
  \bra \phi_{m',\mathbf{k}}^{I,n,l}|\psi_{\mathbf{k},v}^\sigma\ket \bra\frac{\partial\psi_{\mathbf{k},v}^\sigma}{\partial r_{i}}| \phi_{m,\mathbf{k}}^{I,n,l}\ket\Bigg]\,.\nonumber\\
\end{eqnarray}

\begin{figure}[t]
  \begin{center}
    \includegraphics[width=0.9\columnwidth]{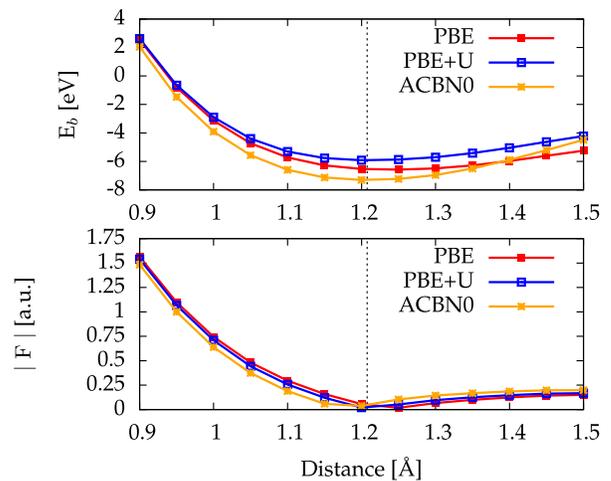}
    \caption{\label{fig:EvsForce} Top panel: Binding energy \textit{versus} atomic distance for the O$_2$ molecule, obtained from PBE (blue curve), PBE+U with $U=4$ eV (red curve), and using the ACBN0 functional (orange curve). Bottom panel: Corresponding absolute value of the atomic force. The vertical line indicated the experimental equilibrium distance of 1.208~\AA .~\cite{huber2013molecular} }
  \end{center}
\end{figure}

We illustrate how our implementation works with a simple, but relevant and widely studied case: the variation of the binding energy and the forces of the O$_2$ dimer in its triplet ground state $^3\Sigma_g^{-}$.
We employed a real-space grid spacing of 0.2 bohr and we fixed the occupations to be the ones of the triplet $^3\Sigma_g^{-}$ state. 
We found (see Fig.~\ref{fig:EvsForce}) that the binding energy and the forces are both consistent in predicting a smaller equilibrium distance when the Hubbard $U$ is added, thus showing the reliability of our implementation.
Interestingly, the self-consistent Hubbard $U$ obtained from the ACBN0 functional predicts an equilibrium distance closer to the experimental distance of 1.208~\AA ~\cite{huber2013molecular} than the use of the PBE functional or the PBE+U with a fixed empirical $U$ of 4~eV, as summarized in Tab.~\ref{tab:bond_length}.

\begin{table}[h]
\begin{ruledtabular}
\begin{tabular}{l|c|c|c|c}
Method & PBE & PBE+U & ACBN0 & Exp.\\
         \hline
Bound length (\AA)  &  1.245 & 1.218 & 1.210 & 1.208\\
Vibration freq. (cm$^{-1}$) & 1552 & 1506 & 1773 & 1580\\
\end{tabular}
\end{ruledtabular}
\caption{\label{tab:bond_length}Calculated values of the equilibrium bond length of O$_2$, in \AA ngstrom, binding energy, in eV, and vibration frequencies, in cm$^{-1}$, obtained for the different methods, compared to the corresponding experimental value.}
\end{table}

\section{Real-time TDDFT+U calculations: optical response functions}
\label{sec:results}
Excited-state properties cannot be obtained from ground-state DFT. Therefore, we extended our DFT+U implementation to also support real-time time-dependent DFT+U. 
Following the usual approach used when extending time-independent functionals to the time-dependent case, we make the adiabatic approximation, which in our case means that
\begin{eqnarray}
 V^{\sigma}_{U}(\mathbf{r},\mathbf{r'},t) &=&\bra \mathbf{r} | \hat{V}^{\sigma}_{U}(t) | \mathbf{r'}\ket  \nonumber\\
 &=& V^{\sigma}_{U}[n(t),\{n_{mm'}^\sigma\}(t),U^{\mathrm{eff}}(t)](\mathbf{r},\mathbf{r'})\,.
 \label{eq:pot_V_U_TD}
\end{eqnarray}
This corresponds to evaluating the potential defined by Eq.~\eqref{eq:pot_V_U} from the density, occupation matrices, and effective $U$ computed from the time-dependent wavefunctions.

Real-time TDDFT+U is an alternative to the more commonly used linear response TDDFT+U, where optical properties are computed from perturbation theory based on the energies and wavefunctions obtained from ground DFT+U calculations.\cite{PhysRevB.62.16392}
In Octopus, we implemented the real-time TDDFT+U method, for both the usual DFT+U scheme (presented in Sec.\ref{sec:DFT_U}) and for the \textit{ab initio} scheme based on the ACBN0 functional (see Sec.~\ref{sec:ACBN0}).

Since the DFT+U potential of Eq.~\eqref{eq:pot_V_U_TD} is a non-local operator, some care is required when extending the DFT+U method to the time-dependent case.
In order to preserve gauge-invariance, we replace Eq.~\eqref{eq:pot_V_U_TD} by
\begin{equation}
 \hat{V}^{\sigma,\mathbf{A}}_{U}(t)|\psi_{n,\mathbf{k}}^{\sigma}(t) \ket  = e^{-i\mathbf{A}(t)\mathbf{\hat{r}}} \hat{V}^\sigma_{U}(t)e^{i\mathbf{A}(t)\mathbf{\hat{r}}}|\psi_{n,\mathbf{k}}^{\sigma}(t) \ket \,,
\end{equation}
where $\mathbf{A}(t)$ denotes the vector potential perturbing the electronic system.
Moreover, the electronic current is modified such that
\begin{eqnarray}
 \mathbf{j}_{\mathrm{DFT+U}}(\mathbf{r},t) = \mathbf{j}_{\mathrm{DFT}}(\mathbf{r},t)  \nonumber\\
    - \frac{1}{2}\sum_{n}\sum_{\mathbf{k}}^{\mathrm{BZ}} w_{\mathbf{k}}f_{n,\mathbf{k}} \bra \psi_{n,\mathbf{k}} | [\mathbf{\hat{r}}, \hat{V}^{\sigma,\mathbf{A}}_{U}(t)]|\psi_{n,\mathbf{k}}  \ket +c.c.\,,
\end{eqnarray}
where $c.c.$ denotes the complex conjugate.

\begin{figure}[t]
  \begin{center}
    \includegraphics[width=0.9\columnwidth]{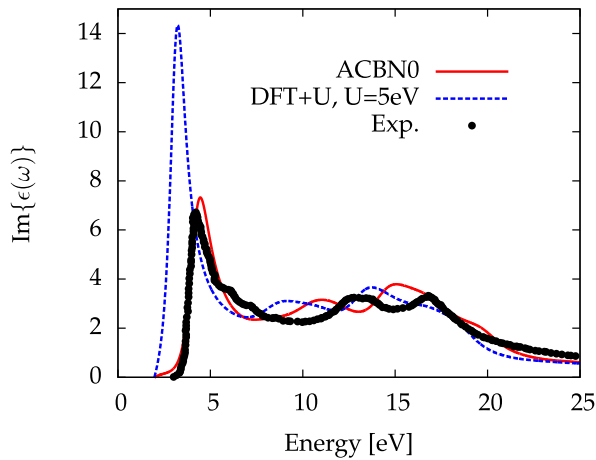}
    \caption{\label{fig:Comp_DFTU_exp} Calculated optical absorption spectra of bulk NiO obtained using the ACBN0 functional (red line), and the usual DFT+U scheme with $U_{\mathrm{eff}}=5$eV (blue dashed line). The experimental data is taken from Ref.~\onlinecite{PhysRevB.2.2182}. A Gaussian broadening of $\eta=0.5$ eV is added to reproduce the experimental broadening.}
  \end{center}
\end{figure}

\begin{figure}[t]
  \begin{center}
    \includegraphics[width=0.9\columnwidth]{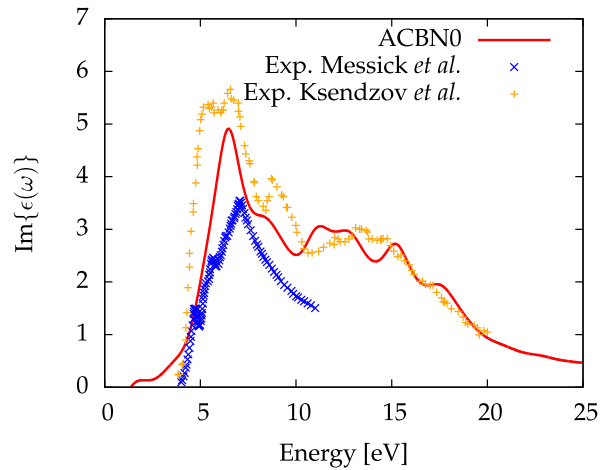}
    \caption{\label{fig:Comp_DFTU_exp_MnO} Calculated optical absorption spectra of bulk MnO obtained using the ACBN0 functional (red line). The experimental data is taken from Messick \textit{et al.}~\cite{Ksendzov,PhysRevB.6.3941} and Ksendzov \textit{et al.}~\cite{rodl_optical_2012}. A Gaussian broadening of $\eta=0.4$ eV is added to reproduce the experimental broadening.}
  \end{center}
\end{figure}

This allows us, for instance, to study linear and non-linear optical properties of transition metal oxides, without relying on perturbation theory. 
To illustrate this, we computed the absorption spectrum of bulk NiO  in its anti-ferromagnetic phase. 
We used a time-step of 0.01 a.u. and propagated the time-dependent Kohn-Sham equations, after a sudden perturbation of the system along the [100] direction, up to 200 a.u. in order to get a converged spectra.
Our results, presented in Fig.~\ref{fig:Comp_DFTU_exp}, show the absorption spectra of NiO obtained using the ACBN0 functional (red line), and using the usual DFT+U scheme with a fixed $U$ of 5 eV (blue dashed line).
We considered localization for both the $3d$ orbitals of the transition metal atoms and the $2p$ orbitals of the O atoms.
We also report in Fig.~\ref{fig:Comp_DFTU_exp_MnO} the optical spectrum of MnO in its anti-ferromagnetic phase.

The optical spectra obtained from the ACBN0 functional yields results in surprisingly good agreement with the experimental spectrum, much better both in pole structure and oscillator strength than the LDA ones. 
This indicates that it is possible to get at the same time good ground state properties, such as the band gap, and good optical spectra using the \textit{ab initio} DFT+U method. 
We found that the excitonic effects are not captured in the absorption spectra of TiO$_2$ and ZnO (not shown).
Adding excitonic contributions to the spectra would be the next step to be considered (as already done for normal semiconductor and insulators using the GW+BSE method ~\cite{RevModPhys.74.601}) and might be achieved by combining the present work with recent developments in describing exciton dynamics using hybrid functionals. See for instance Ref.\onlinecite{PhysRevB.92.081204}.

\section{Conclusion}
\label{sec:conclusions}

We reported an efficient implementation of the DFT+U method and its extension to the time-dependent case in the real-space TDDFT code Octopus.
In the case of ground-state calculations, we showed that our implementation yields results similar to the ones previously reported in the literature.
Moreover, our real-time TDDFT+U implementation is capable of producing good absorption spectra of transition metal oxides, based on the ACBN0 functional, maintaining the relative low cost of the TDDFT+U approach, while being fully \textit{ab initio}.

\acknowledgments
We acknowledge financial support from the European Research Council(ERC-2015-AdG-694097), Grupos Consolidados UPV/EHU (IT578-13), and  European Union's H2020  program under GA no.676580 (NOMAD).
N.T.-D. would like to acknowledge T. Brumme for interesting and fruitful discussions.

\appendix
\section{DFT+U with noncollinear spin}
\label{app_dftu_nc}

Starting from the Hartree-Fock energy of the localized basis for noncollinear spin, we can define $U_{I,n,l}$ and $J_{I,n,l}$~\cite{PhysRevB.85.045132} and obtain
\begin{eqnarray}
  E_{ee}[\{n_{mm'}^{I,m,l,\sigma\sigma'}\}]= \frac{U_{I,n,l}}{2} \sum_{\sigma}\sum_{mm'} n_{mm}^{\sigma\sigma}n_{m'm'}^{-\sigma-\sigma} \nonumber\\
 - \frac{U_{I,n,l}}{2} \sum_{\sigma}\sum_{m} n_{mm}^{\sigma-\sigma}n_{mm}^{-\sigma\sigma}
  - \frac{J_{I,n,l}}{2} \sum_{\sigma}\sum_{m\neq m'} n_{mm}^{\sigma-\sigma}n_{m'm'}^{-\sigma\sigma}\nonumber\\
  + \frac{U_{I,n,l}-J_{I,n,l}}{2} \sum_{\sigma}\sum_{m\neq m'} n_{mm}^{\sigma\sigma}n_{m'm'}^{\sigma\sigma}\,,\nonumber\\
  \label{eq:HF_nc_approx}
\end{eqnarray}
where we assumed for simplicity that the occupation matrices are diagonal.
These occupation matrices are matrices in spin-space and read as
\begin{equation}
 n_{mm'}^{I,m,l,\sigma\sigma'}= \sum_{n}\sum_{\mathbf{k}}^{\mathrm{BZ}}w_{\mathbf{k}}f_{n\mathbf{k}}\bra \psi_{n,\mathbf{k}}^\sigma|\phi_{I,n,l,m}^{\sigma}\ket\bra \phi_{I,n,l,m'}^{\sigma'}|\psi_{n,\mathbf{k}}^{\sigma'}\ket,
\end{equation}
where $f_{n\mathbf{k}}$ is the occupation of the spinor state $|\psi_{n\mathbf{k}}\ket$. Here we consider the most general case, in which the localized basis is described by spinors, but this is not mandatory.

In order to get the DFT+U energy, we need to remove the double-counting part. For this, we use the expression proposed in Ref.~\onlinecite{PhysRevB.80.035121} for the fully-localized limit (FLL)
\begin{equation}
 E_{DC} = \frac{U_{I,n,l}}{2}N(N-1) - \frac{J_{I,n,l}}{2}N(\frac{N}{2}-1) - \frac{J_{I,n,l}}{4}\mathbf{m}\ldotp\mathbf{m}\,,
\end{equation}
where $N=\mathrm{Tr}_{ls}\{n_{mm'}^{\sigma\sigma'}\}$ is the number of electrons in the localized orbitals, i.e. $N = \sum_{\sigma}\sum_m n_{mm}^{\sigma\sigma}$, and $\mathbf{m}$ is the magnetization of the localized subspace. 
In turn, $\mathbf{m}$ is defined as
\begin{equation}
 \mathbf{m} = \mathrm{Tr}_{s}\{\mathbf{\sigma}\ldotp\rho\},\qquad \rho^{\sigma\sigma'}=\mathrm{Tr}_{l}[n_{mm'}^{\sigma\sigma'}]\,,
\end{equation}
with $\mathrm{Tr}_{l}$ the trace over the orbitals, $\sigma_i$ are the Pauli matrices, and $\mathrm{Tr}_{s}$ the trace over spins.

Putting everything together, one obtains that the rotationally-invariant form corresponding to Eq.~\eqref{eq:E_Dudarev} for the noncollinear spins is
\begin{equation}
 E_U =  \frac{U_{I,n,l}-J_{I,n,l}}{2}  \left[ \sum_{\sigma}\sum_{m}n_{mm}^{\sigma\sigma} - \sum_{mm'}\sum_{\sigma\sigma'}n_{mm'}^{\sigma\sigma'}n_{m'm}^{\sigma'\sigma}\right]\,.
\end{equation}

\section{ACBN0 functional with noncollinear spin}
\label{app_acbn0_nc}

In order to derive the expressions needed to compute the ACBN0 functional for noncollinear spin, we first consider the Hartree-Fock energy corresponding to the $\{m\}$ subspace
\begin{eqnarray}
  E_{ee}[\{n_{mm'}^{I,m,l,\sigma\sigma'}\}] =\frac{1}{2} \sum_{\sigma\sigma'}\sum_{\{m\}}  \Big[n_{mm'}^{\sigma\sigma}n_{m''m'''}^{\sigma'\sigma'}(mm'\sigma|m''m'''\sigma') \nonumber\\
  - n_{mm'}^{\sigma'\sigma} n_{m''m'''}^{\sigma\sigma'}(mm'''\sigma|m''m'\sigma')\Big]\,,
  \label{eq:}
\end{eqnarray}
where the Coulomb integrals are denoted using the notation
\begin{eqnarray}
 (mm'\sigma|m''m'''\sigma') = \bra m \sigma, m''\sigma' | V | m'\sigma, m'''\sigma' \ket \nonumber\\
 = \int d^3\mathbf{r_1}\int d^3\mathbf{r_2}  \frac{\phi_{m}^{\sigma *}(\mathbf{r_1})\phi_{m'}^{\sigma}(\mathbf{r_1})\phi_{m''}^{\sigma'*}(\mathbf{r_2})\phi^{\sigma'}_{m'''}(\mathbf{r_2})}{|\mathbf{r_1}-\mathbf{r_2}|}\,.
\end{eqnarray}

Following the idea of Agapito \textit{et al.},~\cite{Agapito_PRX} we introduce the ``renormalized'' occupation number $\bar{N}^{I,n,l}_{\Psi_i}$ (see Eq.~\eqref{eq:notation_Nbar}).\\
We then define the renormalized density matrix as 
\begin{equation}
 \bar{P}^{I,n,l,\sigma\sigma'}_{mm'} = \sum_{n\mathbf{k}}\bar{N}_{\Psi_{n\mathbf{k}}}^{I,n,l} w_{\mathbf{k}}f_{n\mathbf{k}}
 \bra \psi_{n\mathbf{k}}^\sigma|\phi_m^{\sigma}\ket\bra \phi_{m'}^{\sigma'}|\psi_{n\mathbf{k}}^{\sigma'}\ket\,,
\end{equation}
with
  \begin{equation}
  \bar{N}^{I,n,l}_{\psi_{n\mathbf{k}}} = \sum_{\{I\}} \sum_{m} \bra \psi_{n\mathbf{k}} |  \phi_{I,n,l,m}  \ket  \bra \phi_{I,n,l,m} | \psi_{n\mathbf{k}} \ket\,.
 \end{equation}
This definition reduces to the usual expression for the diagonal terms (in spin space), but also includes the screening (from the renormalization) for the off-diagonal terms.\\
Next, we replace $n_{mm'}^{\sigma\sigma'}$ by $\bar{P}_{mm'}^{\sigma\sigma'}$ in the previous expression, and define $\bar{U}$ and $\bar{J}$ by comparing with Eq.~\eqref{eq:HF_nc_approx}.

\begin{widetext}
We obtain that
\begin{equation}
 \bar{U} = \frac{\sum_{\sigma\sigma'}\sum_{\{m\}}  \bar{P}_{mm'}^{\sigma\sigma}\bar{P}_{m''m'''}^{\sigma'\sigma'}(mm'\sigma|m''m'''\sigma') }
 {\sum_{\sigma}\sum_{mm'}N_{m}^{\sigma\sigma}N_{m'}^{-\sigma-\sigma} + \sum_{\sigma}\sum_{m\neq m'} N_{m}^{\sigma\sigma}N_{m'}^{\sigma\sigma}
 - \sum_{\sigma}\sum_{m} N_{m}^{\sigma-\sigma}N_{m}^{-\sigma\sigma}}\,,
\end{equation}
\begin{equation}
 \bar{J} = \frac{\sum_{\sigma\sigma'}\sum_{\{m\}} \bar{P}_{mm'}^{\sigma\sigma'} \bar{P}_{m''m'''}^{\sigma'\sigma}(mm'''\sigma|m''m'\sigma')}{\sum_{\sigma}\sum_{m\neq m'} \left[N_{m}^{\sigma\sigma}N_{m'}^{\sigma\sigma} +  N_{m}^{\sigma-\sigma}N_{m'}^{-\sigma\sigma}\right]}\,,
\end{equation}
with $N_{m}^{\alpha\beta} = n_{mm}^{I,n,l,\alpha,\beta}$.
\end{widetext}

\bibliography{bibliography}

\begin{thebibliography}{46}%
\makeatletter
\providecommand \@ifxundefined [1]{%
 \@ifx{#1\undefined}
}%
\providecommand \@ifnum [1]{%
 \ifnum #1\expandafter \@firstoftwo
 \else \expandafter \@secondoftwo
 \fi
}%
\providecommand \@ifx [1]{%
 \ifx #1\expandafter \@firstoftwo
 \else \expandafter \@secondoftwo
 \fi
}%
\providecommand \natexlab [1]{#1}%
\providecommand \enquote  [1]{``#1''}%
\providecommand \bibnamefont  [1]{#1}%
\providecommand \bibfnamefont [1]{#1}%
\providecommand \citenamefont [1]{#1}%
\providecommand \href@noop [0]{\@secondoftwo}%
\providecommand \href [0]{\begingroup \@sanitize@url \@href}%
\providecommand \@href[1]{\@@startlink{#1}\@@href}%
\providecommand \@@href[1]{\endgroup#1\@@endlink}%
\providecommand \@sanitize@url [0]{\catcode `\\12\catcode `\$12\catcode
  `\&12\catcode `\#12\catcode `\^12\catcode `\_12\catcode `\%12\relax}%
\providecommand \@@startlink[1]{}%
\providecommand \@@endlink[0]{}%
\providecommand \url  [0]{\begingroup\@sanitize@url \@url }%
\providecommand \@url [1]{\endgroup\@href {#1}{\urlprefix }}%
\providecommand \urlprefix  [0]{URL }%
\providecommand \Eprint [0]{\href }%
\providecommand \doibase [0]{http://dx.doi.org/}%
\providecommand \selectlanguage [0]{\@gobble}%
\providecommand \bibinfo  [0]{\@secondoftwo}%
\providecommand \bibfield  [0]{\@secondoftwo}%
\providecommand \translation [1]{[#1]}%
\providecommand \BibitemOpen [0]{}%
\providecommand \bibitemStop [0]{}%
\providecommand \bibitemNoStop [0]{.\EOS\space}%
\providecommand \EOS [0]{\spacefactor3000\relax}%
\providecommand \BibitemShut  [1]{\csname bibitem#1\endcsname}%
\let\auto@bib@innerbib\@empty
\bibitem [{\citenamefont {Anisimov}\ \emph {et~al.}(1997)\citenamefont
  {Anisimov}, \citenamefont {Aryasetiawan},\ and\ \citenamefont
  {Lichtenstein}}]{anisimov1997first}%
  \BibitemOpen
  \bibfield  {author} {\bibinfo {author} {\bibfnamefont {V.~I.}\ \bibnamefont
  {Anisimov}}, \bibinfo {author} {\bibfnamefont {F.}~\bibnamefont
  {Aryasetiawan}}, \ and\ \bibinfo {author} {\bibfnamefont {A.}~\bibnamefont
  {Lichtenstein}},\ }\href@noop {} {\bibfield  {journal} {\bibinfo  {journal}
  {Journal of Physics: Condensed Matter}\ }\textbf {\bibinfo {volume} {9}},\
  \bibinfo {pages} {767} (\bibinfo {year} {1997})}\BibitemShut {NoStop}%
\bibitem [{\citenamefont {Himmetoglu}\ \emph {et~al.}(2014)\citenamefont
  {Himmetoglu}, \citenamefont {Floris}, \citenamefont {Gironcoli},\ and\
  \citenamefont {Cococcioni}}]{himmetoglu2014hubbard}%
  \BibitemOpen
  \bibfield  {author} {\bibinfo {author} {\bibfnamefont {B.}~\bibnamefont
  {Himmetoglu}}, \bibinfo {author} {\bibfnamefont {A.}~\bibnamefont {Floris}},
  \bibinfo {author} {\bibfnamefont {S.}~\bibnamefont {Gironcoli}}, \ and\
  \bibinfo {author} {\bibfnamefont {M.}~\bibnamefont {Cococcioni}},\
  }\href@noop {} {\bibfield  {journal} {\bibinfo  {journal} {International
  Journal of Quantum Chemistry}\ }\textbf {\bibinfo {volume} {114}},\ \bibinfo
  {pages} {14} (\bibinfo {year} {2014})}\BibitemShut {NoStop}%
\bibitem [{\citenamefont {Anisimov}\ \emph {et~al.}(1991)\citenamefont
  {Anisimov}, \citenamefont {Zaanen},\ and\ \citenamefont
  {Andersen}}]{anisimov_band_1991}%
  \BibitemOpen
  \bibfield  {author} {\bibinfo {author} {\bibfnamefont {V.~I.}\ \bibnamefont
  {Anisimov}}, \bibinfo {author} {\bibfnamefont {J.}~\bibnamefont {Zaanen}}, \
  and\ \bibinfo {author} {\bibfnamefont {O.~K.}\ \bibnamefont {Andersen}},\
  }\href@noop {} {\bibfield  {journal} {\bibinfo  {journal} {Physical Review
  B}\ }\textbf {\bibinfo {volume} {44}},\ \bibinfo {pages} {943} (\bibinfo
  {year} {1991})}\BibitemShut {NoStop}%
\bibitem [{\citenamefont {Anisimov}\ \emph {et~al.}(1993)\citenamefont
  {Anisimov}, \citenamefont {Solovyev}, \citenamefont {Korotin}, \citenamefont
  {Czy{\.z}yk},\ and\ \citenamefont
  {Sawatzky}}]{anisimov_density-functional_1993}%
  \BibitemOpen
  \bibfield  {author} {\bibinfo {author} {\bibfnamefont {V.~I.}\ \bibnamefont
  {Anisimov}}, \bibinfo {author} {\bibfnamefont {I.~V.}\ \bibnamefont
  {Solovyev}}, \bibinfo {author} {\bibfnamefont {M.~A.}\ \bibnamefont
  {Korotin}}, \bibinfo {author} {\bibfnamefont {M.~T.}\ \bibnamefont
  {Czy{\.z}yk}}, \ and\ \bibinfo {author} {\bibfnamefont {G.~A.}\ \bibnamefont
  {Sawatzky}},\ }\href@noop {} {\bibfield  {journal} {\bibinfo  {journal}
  {Physical Review B}\ }\textbf {\bibinfo {volume} {48}},\ \bibinfo {pages}
  {16929} (\bibinfo {year} {1993})}\BibitemShut {NoStop}%
\bibitem [{\citenamefont {Liechtenstein}\ \emph {et~al.}(1995)\citenamefont
  {Liechtenstein}, \citenamefont {Anisimov},\ and\ \citenamefont
  {Zaanen}}]{liechtenstein_density-functional_1995}%
  \BibitemOpen
  \bibfield  {author} {\bibinfo {author} {\bibfnamefont {A.~I.}\ \bibnamefont
  {Liechtenstein}}, \bibinfo {author} {\bibfnamefont {V.~I.}\ \bibnamefont
  {Anisimov}}, \ and\ \bibinfo {author} {\bibfnamefont {J.}~\bibnamefont
  {Zaanen}},\ }\href
  {http://journals.aps.org/prb/abstract/10.1103/PhysRevB.52.R5467} {\bibfield
  {journal} {\bibinfo  {journal} {Physical Review B}\ }\textbf {\bibinfo
  {volume} {52}},\ \bibinfo {pages} {R5467} (\bibinfo {year}
  {1995})}\BibitemShut {NoStop}%
\bibitem [{\citenamefont {Perdew}\ and\ \citenamefont
  {Wang}(1992)}]{PhysRevB.45.13244}%
  \BibitemOpen
  \bibfield  {author} {\bibinfo {author} {\bibfnamefont {J.~P.}\ \bibnamefont
  {Perdew}}\ and\ \bibinfo {author} {\bibfnamefont {Y.}~\bibnamefont {Wang}},\
  }\href {\doibase 10.1103/PhysRevB.45.13244} {\bibfield  {journal} {\bibinfo
  {journal} {Phys. Rev. B}\ }\textbf {\bibinfo {volume} {45}},\ \bibinfo
  {pages} {13244} (\bibinfo {year} {1992})}\BibitemShut {NoStop}%
\bibitem [{\citenamefont {Perdew}\ \emph {et~al.}(1996)\citenamefont {Perdew},
  \citenamefont {Burke},\ and\ \citenamefont
  {Ernzerhof}}]{PhysRevLett.77.3865}%
  \BibitemOpen
  \bibfield  {author} {\bibinfo {author} {\bibfnamefont {J.~P.}\ \bibnamefont
  {Perdew}}, \bibinfo {author} {\bibfnamefont {K.}~\bibnamefont {Burke}}, \
  and\ \bibinfo {author} {\bibfnamefont {M.}~\bibnamefont {Ernzerhof}},\ }\href
  {\doibase 10.1103/PhysRevLett.77.3865} {\bibfield  {journal} {\bibinfo
  {journal} {Phys. Rev. Lett.}\ }\textbf {\bibinfo {volume} {77}},\ \bibinfo
  {pages} {3865} (\bibinfo {year} {1996})}\BibitemShut {NoStop}%
\bibitem [{\citenamefont {Metzner}\ and\ \citenamefont
  {Vollhardt}(1989)}]{PhysRevLett.62.324}%
  \BibitemOpen
  \bibfield  {author} {\bibinfo {author} {\bibfnamefont {W.}~\bibnamefont
  {Metzner}}\ and\ \bibinfo {author} {\bibfnamefont {D.}~\bibnamefont
  {Vollhardt}},\ }\href {\doibase 10.1103/PhysRevLett.62.324} {\bibfield
  {journal} {\bibinfo  {journal} {Phys. Rev. Lett.}\ }\textbf {\bibinfo
  {volume} {62}},\ \bibinfo {pages} {324} (\bibinfo {year} {1989})}\BibitemShut
  {NoStop}%
\bibitem [{\citenamefont {Georges}\ and\ \citenamefont
  {Kotliar}(1992)}]{PhysRevB.45.6479}%
  \BibitemOpen
  \bibfield  {author} {\bibinfo {author} {\bibfnamefont {A.}~\bibnamefont
  {Georges}}\ and\ \bibinfo {author} {\bibfnamefont {G.}~\bibnamefont
  {Kotliar}},\ }\href {\doibase 10.1103/PhysRevB.45.6479} {\bibfield  {journal}
  {\bibinfo  {journal} {Phys. Rev. B}\ }\textbf {\bibinfo {volume} {45}},\
  \bibinfo {pages} {6479} (\bibinfo {year} {1992})}\BibitemShut {NoStop}%
\bibitem [{\citenamefont {Kotliar}\ \emph {et~al.}(2006)\citenamefont
  {Kotliar}, \citenamefont {Savrasov}, \citenamefont {Haule}, \citenamefont
  {Oudovenko}, \citenamefont {Parcollet},\ and\ \citenamefont
  {Marianetti}}]{RevModPhys.78.865}%
  \BibitemOpen
  \bibfield  {author} {\bibinfo {author} {\bibfnamefont {G.}~\bibnamefont
  {Kotliar}}, \bibinfo {author} {\bibfnamefont {S.~Y.}\ \bibnamefont
  {Savrasov}}, \bibinfo {author} {\bibfnamefont {K.}~\bibnamefont {Haule}},
  \bibinfo {author} {\bibfnamefont {V.~S.}\ \bibnamefont {Oudovenko}}, \bibinfo
  {author} {\bibfnamefont {O.}~\bibnamefont {Parcollet}}, \ and\ \bibinfo
  {author} {\bibfnamefont {C.~A.}\ \bibnamefont {Marianetti}},\ }\href
  {\doibase 10.1103/RevModPhys.78.865} {\bibfield  {journal} {\bibinfo
  {journal} {Rev. Mod. Phys.}\ }\textbf {\bibinfo {volume} {78}},\ \bibinfo
  {pages} {865} (\bibinfo {year} {2006})}\BibitemShut {NoStop}%
\bibitem [{\citenamefont {Cococcioni}\ and\ \citenamefont
  {de~Gironcoli}(2005)}]{PhysRevB.71.035105}%
  \BibitemOpen
  \bibfield  {author} {\bibinfo {author} {\bibfnamefont {M.}~\bibnamefont
  {Cococcioni}}\ and\ \bibinfo {author} {\bibfnamefont {S.}~\bibnamefont
  {de~Gironcoli}},\ }\href {\doibase 10.1103/PhysRevB.71.035105} {\bibfield
  {journal} {\bibinfo  {journal} {Phys. Rev. B}\ }\textbf {\bibinfo {volume}
  {71}},\ \bibinfo {pages} {035105} (\bibinfo {year} {2005})}\BibitemShut
  {NoStop}%
\bibitem [{\citenamefont {Miyake}\ and\ \citenamefont
  {Aryasetiawan}(2008)}]{miyake_screened_2008}%
  \BibitemOpen
  \bibfield  {author} {\bibinfo {author} {\bibfnamefont {T.}~\bibnamefont
  {Miyake}}\ and\ \bibinfo {author} {\bibfnamefont {F.}~\bibnamefont
  {Aryasetiawan}},\ }\href {\doibase 10.1103/PhysRevB.77.085122} {\bibfield
  {journal} {\bibinfo  {journal} {Phys. Rev. B}\ }\textbf {\bibinfo {volume}
  {77}},\ \bibinfo {pages} {085122} (\bibinfo {year} {2008})}\BibitemShut
  {NoStop}%
\bibitem [{\citenamefont {Aichhorn}\ \emph {et~al.}(2009)\citenamefont
  {Aichhorn}, \citenamefont {Pourovskii}, \citenamefont {Vildosola},
  \citenamefont {Ferrero}, \citenamefont {Parcollet}, \citenamefont {Miyake},
  \citenamefont {Georges},\ and\ \citenamefont
  {Biermann}}]{aichhorn_dynamical_2009}%
  \BibitemOpen
  \bibfield  {author} {\bibinfo {author} {\bibfnamefont {M.}~\bibnamefont
  {Aichhorn}}, \bibinfo {author} {\bibfnamefont {L.}~\bibnamefont
  {Pourovskii}}, \bibinfo {author} {\bibfnamefont {V.}~\bibnamefont
  {Vildosola}}, \bibinfo {author} {\bibfnamefont {M.}~\bibnamefont {Ferrero}},
  \bibinfo {author} {\bibfnamefont {O.}~\bibnamefont {Parcollet}}, \bibinfo
  {author} {\bibfnamefont {T.}~\bibnamefont {Miyake}}, \bibinfo {author}
  {\bibfnamefont {A.}~\bibnamefont {Georges}}, \ and\ \bibinfo {author}
  {\bibfnamefont {S.}~\bibnamefont {Biermann}},\ }\href {\doibase
  10.1103/PhysRevB.80.085101} {\bibfield  {journal} {\bibinfo  {journal} {Phys.
  Rev. B}\ }\textbf {\bibinfo {volume} {80}},\ \bibinfo {pages} {085101}
  (\bibinfo {year} {2009})}\BibitemShut {NoStop}%
\bibitem [{\citenamefont {Miyake}\ \emph {et~al.}(2009)\citenamefont {Miyake},
  \citenamefont {Aryasetiawan},\ and\ \citenamefont {Imada}}]{miyake_ab_2009}%
  \BibitemOpen
  \bibfield  {author} {\bibinfo {author} {\bibfnamefont {T.}~\bibnamefont
  {Miyake}}, \bibinfo {author} {\bibfnamefont {F.}~\bibnamefont
  {Aryasetiawan}}, \ and\ \bibinfo {author} {\bibfnamefont {M.}~\bibnamefont
  {Imada}},\ }\href {\doibase 10.1103/PhysRevB.80.155134} {\bibfield  {journal}
  {\bibinfo  {journal} {Phys. Rev. B}\ }\textbf {\bibinfo {volume} {80}},\
  \bibinfo {pages} {155134} (\bibinfo {year} {2009})}\BibitemShut {NoStop}%
\bibitem [{\citenamefont {Aichhorn}\ \emph {et~al.}(2010)\citenamefont
  {Aichhorn}, \citenamefont {Biermann}, \citenamefont {Miyake}, \citenamefont
  {Georges},\ and\ \citenamefont {Imada}}]{aichhorn_theoretical_2010}%
  \BibitemOpen
  \bibfield  {author} {\bibinfo {author} {\bibfnamefont {M.}~\bibnamefont
  {Aichhorn}}, \bibinfo {author} {\bibfnamefont {S.}~\bibnamefont {Biermann}},
  \bibinfo {author} {\bibfnamefont {T.}~\bibnamefont {Miyake}}, \bibinfo
  {author} {\bibfnamefont {A.}~\bibnamefont {Georges}}, \ and\ \bibinfo
  {author} {\bibfnamefont {M.}~\bibnamefont {Imada}},\ }\href {\doibase
  10.1103/PhysRevB.82.064504} {\bibfield  {journal} {\bibinfo  {journal} {Phys.
  Rev. B}\ }\textbf {\bibinfo {volume} {82}},\ \bibinfo {pages} {064504}
  (\bibinfo {year} {2010})}\BibitemShut {NoStop}%
\bibitem [{\citenamefont {Agapito}\ \emph {et~al.}(2015)\citenamefont
  {Agapito}, \citenamefont {Curtarolo},\ and\ \citenamefont
  {Buongiorno~Nardelli}}]{Agapito_PRX}%
  \BibitemOpen
  \bibfield  {author} {\bibinfo {author} {\bibfnamefont {L.~A.}\ \bibnamefont
  {Agapito}}, \bibinfo {author} {\bibfnamefont {S.}~\bibnamefont {Curtarolo}},
  \ and\ \bibinfo {author} {\bibfnamefont {M.}~\bibnamefont
  {Buongiorno~Nardelli}},\ }\href {\doibase 10.1103/PhysRevX.5.011006}
  {\bibfield  {journal} {\bibinfo  {journal} {Phys. Rev. X}\ }\textbf {\bibinfo
  {volume} {5}},\ \bibinfo {pages} {011006} (\bibinfo {year}
  {2015})}\BibitemShut {NoStop}%
\bibitem [{\citenamefont {Mosey}\ and\ \citenamefont
  {Carter}(2007)}]{PhysRevB.76.155123}%
  \BibitemOpen
  \bibfield  {author} {\bibinfo {author} {\bibfnamefont {N.~J.}\ \bibnamefont
  {Mosey}}\ and\ \bibinfo {author} {\bibfnamefont {E.~A.}\ \bibnamefont
  {Carter}},\ }\href {\doibase 10.1103/PhysRevB.76.155123} {\bibfield
  {journal} {\bibinfo  {journal} {Phys. Rev. B}\ }\textbf {\bibinfo {volume}
  {76}},\ \bibinfo {pages} {155123} (\bibinfo {year} {2007})}\BibitemShut
  {NoStop}%
\bibitem [{\citenamefont {Kulik}\ \emph {et~al.}(2006)\citenamefont {Kulik},
  \citenamefont {Cococcioni}, \citenamefont {Scherlis},\ and\ \citenamefont
  {Marzari}}]{PhysRevLett.97.103001}%
  \BibitemOpen
  \bibfield  {author} {\bibinfo {author} {\bibfnamefont {H.~J.}\ \bibnamefont
  {Kulik}}, \bibinfo {author} {\bibfnamefont {M.}~\bibnamefont {Cococcioni}},
  \bibinfo {author} {\bibfnamefont {D.~A.}\ \bibnamefont {Scherlis}}, \ and\
  \bibinfo {author} {\bibfnamefont {N.}~\bibnamefont {Marzari}},\ }\href
  {\doibase 10.1103/PhysRevLett.97.103001} {\bibfield  {journal} {\bibinfo
  {journal} {Phys. Rev. Lett.}\ }\textbf {\bibinfo {volume} {97}},\ \bibinfo
  {pages} {103001} (\bibinfo {year} {2006})}\BibitemShut {NoStop}%
\bibitem [{\citenamefont {Marques}\ \emph {et~al.}(2003)\citenamefont
  {Marques}, \citenamefont {Castro}, \citenamefont {Bertsch},\ and\
  \citenamefont {Rubio}}]{MARQUES200360}%
  \BibitemOpen
  \bibfield  {author} {\bibinfo {author} {\bibfnamefont {M.~A.}\ \bibnamefont
  {Marques}}, \bibinfo {author} {\bibfnamefont {A.}~\bibnamefont {Castro}},
  \bibinfo {author} {\bibfnamefont {G.~F.}\ \bibnamefont {Bertsch}}, \ and\
  \bibinfo {author} {\bibfnamefont {A.}~\bibnamefont {Rubio}},\ }\href
  {\doibase https://doi.org/10.1016/S0010-4655(02)00686-0} {\bibfield
  {journal} {\bibinfo  {journal} {Computer Physics Communications}\ }\textbf
  {\bibinfo {volume} {151}},\ \bibinfo {pages} {60 } (\bibinfo {year}
  {2003})}\BibitemShut {NoStop}%
\bibitem [{\citenamefont {Castro}\ \emph {et~al.}(2006)\citenamefont {Castro},
  \citenamefont {Appel}, \citenamefont {Oliveira}, \citenamefont {Rozzi},
  \citenamefont {Andrade}, \citenamefont {Lorenzen}, \citenamefont {Marques},
  \citenamefont {Gross},\ and\ \citenamefont {Rubio}}]{castro_octopus:_2006}%
  \BibitemOpen
  \bibfield  {author} {\bibinfo {author} {\bibfnamefont {A.}~\bibnamefont
  {Castro}}, \bibinfo {author} {\bibfnamefont {H.}~\bibnamefont {Appel}},
  \bibinfo {author} {\bibfnamefont {M.}~\bibnamefont {Oliveira}}, \bibinfo
  {author} {\bibfnamefont {C.~A.}\ \bibnamefont {Rozzi}}, \bibinfo {author}
  {\bibfnamefont {X.}~\bibnamefont {Andrade}}, \bibinfo {author} {\bibfnamefont
  {F.}~\bibnamefont {Lorenzen}}, \bibinfo {author} {\bibfnamefont {M.~A.~L.}\
  \bibnamefont {Marques}}, \bibinfo {author} {\bibfnamefont {E.~K.~U.}\
  \bibnamefont {Gross}}, \ and\ \bibinfo {author} {\bibfnamefont
  {A.}~\bibnamefont {Rubio}},\ }\href {\doibase 10.1002/pssb.200642067}
  {\bibfield  {journal} {\bibinfo  {journal} {physica status solidi (b)}\
  }\textbf {\bibinfo {volume} {243}},\ \bibinfo {pages} {2465} (\bibinfo {year}
  {2006})}\BibitemShut {NoStop}%
\bibitem [{\citenamefont {Andrade}\ \emph {et~al.}(2015)\citenamefont
  {Andrade}, \citenamefont {Strubbe}, \citenamefont {De~Giovannini},
  \citenamefont {Larsen}, \citenamefont {Oliveira}, \citenamefont
  {Alberdi-Rodriguez}, \citenamefont {Varas}, \citenamefont {Theophilou},
  \citenamefont {Helbig}, \citenamefont {Verstraete}, \citenamefont {Stella},
  \citenamefont {Nogueira}, \citenamefont {Aspuru-Guzik}, \citenamefont
  {Castro}, \citenamefont {Marques},\ and\ \citenamefont
  {Rubio}}]{andrade_real-space_2015}%
  \BibitemOpen
  \bibfield  {author} {\bibinfo {author} {\bibfnamefont {X.}~\bibnamefont
  {Andrade}}, \bibinfo {author} {\bibfnamefont {D.}~\bibnamefont {Strubbe}},
  \bibinfo {author} {\bibfnamefont {U.}~\bibnamefont {De~Giovannini}}, \bibinfo
  {author} {\bibfnamefont {A.~H.}\ \bibnamefont {Larsen}}, \bibinfo {author}
  {\bibfnamefont {M.~J.~T.}\ \bibnamefont {Oliveira}}, \bibinfo {author}
  {\bibfnamefont {J.}~\bibnamefont {Alberdi-Rodriguez}}, \bibinfo {author}
  {\bibfnamefont {A.}~\bibnamefont {Varas}}, \bibinfo {author} {\bibfnamefont
  {I.}~\bibnamefont {Theophilou}}, \bibinfo {author} {\bibfnamefont
  {N.}~\bibnamefont {Helbig}}, \bibinfo {author} {\bibfnamefont {M.~J.}\
  \bibnamefont {Verstraete}}, \bibinfo {author} {\bibfnamefont
  {L.}~\bibnamefont {Stella}}, \bibinfo {author} {\bibfnamefont
  {F.}~\bibnamefont {Nogueira}}, \bibinfo {author} {\bibfnamefont
  {A.}~\bibnamefont {Aspuru-Guzik}}, \bibinfo {author} {\bibfnamefont
  {A.}~\bibnamefont {Castro}}, \bibinfo {author} {\bibfnamefont {M.~A.~L.}\
  \bibnamefont {Marques}}, \ and\ \bibinfo {author} {\bibfnamefont
  {A.}~\bibnamefont {Rubio}},\ }\href {\doibase 10.1039/C5CP00351B} {\bibfield
  {journal} {\bibinfo  {journal} {Phys. Chem. Chem. Phys.}\ }\textbf {\bibinfo
  {volume} {17}},\ \bibinfo {pages} {31371} (\bibinfo {year}
  {2015})}\BibitemShut {NoStop}%
\bibitem [{\citenamefont {Haule}(2015)}]{PhysRevLett.115.196403}%
  \BibitemOpen
  \bibfield  {author} {\bibinfo {author} {\bibfnamefont {K.}~\bibnamefont
  {Haule}},\ }\href {\doibase 10.1103/PhysRevLett.115.196403} {\bibfield
  {journal} {\bibinfo  {journal} {Phys. Rev. Lett.}\ }\textbf {\bibinfo
  {volume} {115}},\ \bibinfo {pages} {196403} (\bibinfo {year}
  {2015})}\BibitemShut {NoStop}%
\bibitem [{\citenamefont {Dudarev}\ \emph {et~al.}(1998)\citenamefont
  {Dudarev}, \citenamefont {Botton}, \citenamefont {Savrasov}, \citenamefont
  {Humphreys},\ and\ \citenamefont {Sutton}}]{PhysRevB.57.1505}%
  \BibitemOpen
  \bibfield  {author} {\bibinfo {author} {\bibfnamefont {S.~L.}\ \bibnamefont
  {Dudarev}}, \bibinfo {author} {\bibfnamefont {G.~A.}\ \bibnamefont {Botton}},
  \bibinfo {author} {\bibfnamefont {S.~Y.}\ \bibnamefont {Savrasov}}, \bibinfo
  {author} {\bibfnamefont {C.~J.}\ \bibnamefont {Humphreys}}, \ and\ \bibinfo
  {author} {\bibfnamefont {A.~P.}\ \bibnamefont {Sutton}},\ }\href {\doibase
  10.1103/PhysRevB.57.1505} {\bibfield  {journal} {\bibinfo  {journal} {Phys.
  Rev. B}\ }\textbf {\bibinfo {volume} {57}},\ \bibinfo {pages} {1505}
  (\bibinfo {year} {1998})}\BibitemShut {NoStop}%
\bibitem [{\citenamefont {Shick}\ \emph {et~al.}(1999)\citenamefont {Shick},
  \citenamefont {Liechtenstein},\ and\ \citenamefont
  {Pickett}}]{PhysRevB.60.10763}%
  \BibitemOpen
  \bibfield  {author} {\bibinfo {author} {\bibfnamefont {A.~B.}\ \bibnamefont
  {Shick}}, \bibinfo {author} {\bibfnamefont {A.~I.}\ \bibnamefont
  {Liechtenstein}}, \ and\ \bibinfo {author} {\bibfnamefont {W.~E.}\
  \bibnamefont {Pickett}},\ }\href {\doibase 10.1103/PhysRevB.60.10763}
  {\bibfield  {journal} {\bibinfo  {journal} {Phys. Rev. B}\ }\textbf {\bibinfo
  {volume} {60}},\ \bibinfo {pages} {10763} (\bibinfo {year}
  {1999})}\BibitemShut {NoStop}%
\bibitem [{\citenamefont {Perdew}\ and\ \citenamefont
  {Zunger}(1981)}]{PhysRevB.23.5048}%
  \BibitemOpen
  \bibfield  {author} {\bibinfo {author} {\bibfnamefont {J.~P.}\ \bibnamefont
  {Perdew}}\ and\ \bibinfo {author} {\bibfnamefont {A.}~\bibnamefont
  {Zunger}},\ }\href {\doibase 10.1103/PhysRevB.23.5048} {\bibfield  {journal}
  {\bibinfo  {journal} {Phys. Rev. B}\ }\textbf {\bibinfo {volume} {23}},\
  \bibinfo {pages} {5048} (\bibinfo {year} {1981})}\BibitemShut {NoStop}%
\bibitem [{\citenamefont {Janak}(1978)}]{PhysRevB.18.7165}%
  \BibitemOpen
  \bibfield  {author} {\bibinfo {author} {\bibfnamefont {J.~F.}\ \bibnamefont
  {Janak}},\ }\href {\doibase 10.1103/PhysRevB.18.7165} {\bibfield  {journal}
  {\bibinfo  {journal} {Phys. Rev. B}\ }\textbf {\bibinfo {volume} {18}},\
  \bibinfo {pages} {7165} (\bibinfo {year} {1978})}\BibitemShut {NoStop}%
\bibitem [{\citenamefont {Sclauzero}\ and\ \citenamefont
  {Dal~Corso}(2013)}]{PhysRevB.87.085108}%
  \BibitemOpen
  \bibfield  {author} {\bibinfo {author} {\bibfnamefont {G.}~\bibnamefont
  {Sclauzero}}\ and\ \bibinfo {author} {\bibfnamefont {A.}~\bibnamefont
  {Dal~Corso}},\ }\href {\doibase 10.1103/PhysRevB.87.085108} {\bibfield
  {journal} {\bibinfo  {journal} {Phys. Rev. B}\ }\textbf {\bibinfo {volume}
  {87}},\ \bibinfo {pages} {085108} (\bibinfo {year} {2013})}\BibitemShut
  {NoStop}%
\bibitem [{\citenamefont {Bengone}\ \emph {et~al.}(2000)\citenamefont
  {Bengone}, \citenamefont {Alouani}, \citenamefont {Bl\"ochl},\ and\
  \citenamefont {Hugel}}]{PhysRevB.62.16392}%
  \BibitemOpen
  \bibfield  {author} {\bibinfo {author} {\bibfnamefont {O.}~\bibnamefont
  {Bengone}}, \bibinfo {author} {\bibfnamefont {M.}~\bibnamefont {Alouani}},
  \bibinfo {author} {\bibfnamefont {P.}~\bibnamefont {Bl\"ochl}}, \ and\
  \bibinfo {author} {\bibfnamefont {J.}~\bibnamefont {Hugel}},\ }\href
  {\doibase 10.1103/PhysRevB.62.16392} {\bibfield  {journal} {\bibinfo
  {journal} {Phys. Rev. B}\ }\textbf {\bibinfo {volume} {62}},\ \bibinfo
  {pages} {16392} (\bibinfo {year} {2000})}\BibitemShut {NoStop}%
\bibitem [{\citenamefont {Amadon}\ \emph {et~al.}(2008)\citenamefont {Amadon},
  \citenamefont {Jollet},\ and\ \citenamefont {Torrent}}]{amadon__2008}%
  \BibitemOpen
  \bibfield  {author} {\bibinfo {author} {\bibfnamefont {B.}~\bibnamefont
  {Amadon}}, \bibinfo {author} {\bibfnamefont {F.}~\bibnamefont {Jollet}}, \
  and\ \bibinfo {author} {\bibfnamefont {M.}~\bibnamefont {Torrent}},\ }\href
  {\doibase 10.1103/PhysRevB.77.155104} {\bibfield  {journal} {\bibinfo
  {journal} {Phys. Rev. B}\ }\textbf {\bibinfo {volume} {77}},\ \bibinfo
  {pages} {155104} (\bibinfo {year} {2008})}\BibitemShut {NoStop}%
\bibitem [{\citenamefont {Mori-S\'anchez}\ \emph {et~al.}(2008)\citenamefont
  {Mori-S\'anchez}, \citenamefont {Cohen},\ and\ \citenamefont
  {Yang}}]{PhysRevLett.100.146401}%
  \BibitemOpen
  \bibfield  {author} {\bibinfo {author} {\bibfnamefont {P.}~\bibnamefont
  {Mori-S\'anchez}}, \bibinfo {author} {\bibfnamefont {A.~J.}\ \bibnamefont
  {Cohen}}, \ and\ \bibinfo {author} {\bibfnamefont {W.}~\bibnamefont {Yang}},\
  }\href {\doibase 10.1103/PhysRevLett.100.146401} {\bibfield  {journal}
  {\bibinfo  {journal} {Phys. Rev. Lett.}\ }\textbf {\bibinfo {volume} {100}},\
  \bibinfo {pages} {146401} (\bibinfo {year} {2008})}\BibitemShut {NoStop}%
\bibitem [{\citenamefont {Sharp}\ and\ \citenamefont
  {Horton}(1953)}]{PhysRev.90.317}%
  \BibitemOpen
  \bibfield  {author} {\bibinfo {author} {\bibfnamefont {R.~T.}\ \bibnamefont
  {Sharp}}\ and\ \bibinfo {author} {\bibfnamefont {G.~K.}\ \bibnamefont
  {Horton}},\ }\href {\doibase 10.1103/PhysRev.90.317} {\bibfield  {journal}
  {\bibinfo  {journal} {Phys. Rev.}\ }\textbf {\bibinfo {volume} {90}},\
  \bibinfo {pages} {317} (\bibinfo {year} {1953})}\BibitemShut {NoStop}%
\bibitem [{\citenamefont {Talman}\ and\ \citenamefont
  {Shadwick}(1976)}]{PhysRevA.14.36}%
  \BibitemOpen
  \bibfield  {author} {\bibinfo {author} {\bibfnamefont {J.~D.}\ \bibnamefont
  {Talman}}\ and\ \bibinfo {author} {\bibfnamefont {W.~F.}\ \bibnamefont
  {Shadwick}},\ }\href {\doibase 10.1103/PhysRevA.14.36} {\bibfield  {journal}
  {\bibinfo  {journal} {Phys. Rev. A}\ }\textbf {\bibinfo {volume} {14}},\
  \bibinfo {pages} {36} (\bibinfo {year} {1976})}\BibitemShut {NoStop}%
\bibitem [{\citenamefont {K\"ummel}\ and\ \citenamefont
  {Perdew}(2003)}]{PhysRevLett.90.043004}%
  \BibitemOpen
  \bibfield  {author} {\bibinfo {author} {\bibfnamefont {S.}~\bibnamefont
  {K\"ummel}}\ and\ \bibinfo {author} {\bibfnamefont {J.~P.}\ \bibnamefont
  {Perdew}},\ }\href {\doibase 10.1103/PhysRevLett.90.043004} {\bibfield
  {journal} {\bibinfo  {journal} {Phys. Rev. Lett.}\ }\textbf {\bibinfo
  {volume} {90}},\ \bibinfo {pages} {043004} (\bibinfo {year}
  {2003})}\BibitemShut {NoStop}%
\bibitem [{\citenamefont {Abrahams}\ and\ \citenamefont
  {Bernstein}(1971)}]{doi:10.1063/1.1676569}%
  \BibitemOpen
  \bibfield  {author} {\bibinfo {author} {\bibfnamefont {S.~C.}\ \bibnamefont
  {Abrahams}}\ and\ \bibinfo {author} {\bibfnamefont {J.~L.}\ \bibnamefont
  {Bernstein}},\ }\href {\doibase 10.1063/1.1676569} {\bibfield  {journal}
  {\bibinfo  {journal} {The Journal of Chemical Physics}\ }\textbf {\bibinfo
  {volume} {55}},\ \bibinfo {pages} {3206} (\bibinfo {year} {1971})},\ \Eprint
  {http://arxiv.org/abs/http://dx.doi.org/10.1063/1.1676569}
  {http://dx.doi.org/10.1063/1.1676569} \BibitemShut {NoStop}%
\bibitem [{\citenamefont {Hartwigsen}\ \emph {et~al.}(1998)\citenamefont
  {Hartwigsen}, \citenamefont {Goedecker},\ and\ \citenamefont
  {Hutter}}]{HGH_pseudos}%
  \BibitemOpen
  \bibfield  {author} {\bibinfo {author} {\bibfnamefont {C.}~\bibnamefont
  {Hartwigsen}}, \bibinfo {author} {\bibfnamefont {S.}~\bibnamefont
  {Goedecker}}, \ and\ \bibinfo {author} {\bibfnamefont {J.}~\bibnamefont
  {Hutter}},\ }\href {\doibase 10.1103/PhysRevB.58.3641} {\bibfield  {journal}
  {\bibinfo  {journal} {Phys. Rev. B}\ }\textbf {\bibinfo {volume} {58}},\
  \bibinfo {pages} {3641} (\bibinfo {year} {1998})}\BibitemShut {NoStop}%
\bibitem [{\citenamefont {Wyckoff}(1963)}]{wyckoff1963crystal}%
  \BibitemOpen
  \bibfield  {author} {\bibinfo {author} {\bibfnamefont {R.}~\bibnamefont
  {Wyckoff}},\ }\href {https://books.google.de/books?id=YOGUnQEACAAJ} {\emph
  {\bibinfo {title} {Crystal Structures}}},\ \bibinfo {series} {CRYSTAL
  STRUCTURES, 2ND EDITION Series}\ No.\ \bibinfo {number} {Bd. 4}\ (\bibinfo
  {publisher} {Interscience},\ \bibinfo {year} {1963})\BibitemShut {NoStop}%
\bibitem [{\citenamefont {Aguilera}\ \emph {et~al.}(2013)\citenamefont
  {Aguilera}, \citenamefont {Friedrich}, \citenamefont {Bihlmayer},\ and\
  \citenamefont {Bl\"ugel}}]{PhysRevB.88.045206}%
  \BibitemOpen
  \bibfield  {author} {\bibinfo {author} {\bibfnamefont {I.}~\bibnamefont
  {Aguilera}}, \bibinfo {author} {\bibfnamefont {C.}~\bibnamefont {Friedrich}},
  \bibinfo {author} {\bibfnamefont {G.}~\bibnamefont {Bihlmayer}}, \ and\
  \bibinfo {author} {\bibfnamefont {S.}~\bibnamefont {Bl\"ugel}},\ }\href
  {\doibase 10.1103/PhysRevB.88.045206} {\bibfield  {journal} {\bibinfo
  {journal} {Phys. Rev. B}\ }\textbf {\bibinfo {volume} {88}},\ \bibinfo
  {pages} {045206} (\bibinfo {year} {2013})}\BibitemShut {NoStop}%
\bibitem [{\citenamefont {Huber}(2013)}]{huber2013molecular}%
  \BibitemOpen
  \bibfield  {author} {\bibinfo {author} {\bibfnamefont {K.-P.}\ \bibnamefont
  {Huber}},\ }\href@noop {} {\emph {\bibinfo {title} {Molecular spectra and
  molecular structure: IV. Constants of diatomic molecules}}}\ (\bibinfo
  {publisher} {Springer Science \& Business Media},\ \bibinfo {year}
  {2013})\BibitemShut {NoStop}%
\bibitem [{\citenamefont {Powell}\ and\ \citenamefont
  {Spicer}(1970)}]{PhysRevB.2.2182}%
  \BibitemOpen
  \bibfield  {author} {\bibinfo {author} {\bibfnamefont {R.~J.}\ \bibnamefont
  {Powell}}\ and\ \bibinfo {author} {\bibfnamefont {W.~E.}\ \bibnamefont
  {Spicer}},\ }\href {\doibase 10.1103/PhysRevB.2.2182} {\bibfield  {journal}
  {\bibinfo  {journal} {Phys. Rev. B}\ }\textbf {\bibinfo {volume} {2}},\
  \bibinfo {pages} {2182} (\bibinfo {year} {1970})}\BibitemShut {NoStop}%
\bibitem [{\citenamefont {Ksendzov}\ \emph {et~al.}(1976)\citenamefont
  {Ksendzov}, \citenamefont {Koroboa}, \citenamefont {K.},\ and\ \citenamefont
  {P.}}]{Ksendzov}%
  \BibitemOpen
  \bibfield  {author} {\bibinfo {author} {\bibfnamefont {Y.~M.}\ \bibnamefont
  {Ksendzov}}, \bibinfo {author} {\bibfnamefont {I.~L.}\ \bibnamefont
  {Koroboa}}, \bibinfo {author} {\bibfnamefont {S.~K.}\ \bibnamefont {K.}}, \
  and\ \bibinfo {author} {\bibfnamefont {S.~G.}\ \bibnamefont {P.}},\
  }\href@noop {} {\bibfield  {journal} {\bibinfo  {journal} {Fiz. Tverd. Tela
  (Leningrad)}\ }\textbf {\bibinfo {volume} {18}},\ \bibinfo {pages} {173}
  (\bibinfo {year} {1976})}\BibitemShut {NoStop}%
\bibitem [{\citenamefont {Messick}\ \emph {et~al.}(1972)\citenamefont
  {Messick}, \citenamefont {Walker},\ and\ \citenamefont
  {Glosser}}]{PhysRevB.6.3941}%
  \BibitemOpen
  \bibfield  {author} {\bibinfo {author} {\bibfnamefont {L.}~\bibnamefont
  {Messick}}, \bibinfo {author} {\bibfnamefont {W.~C.}\ \bibnamefont {Walker}},
  \ and\ \bibinfo {author} {\bibfnamefont {R.}~\bibnamefont {Glosser}},\ }\href
  {\doibase 10.1103/PhysRevB.6.3941} {\bibfield  {journal} {\bibinfo  {journal}
  {Phys. Rev. B}\ }\textbf {\bibinfo {volume} {6}},\ \bibinfo {pages} {3941}
  (\bibinfo {year} {1972})}\BibitemShut {NoStop}%
\bibitem [{\citenamefont {R\"odl}\ and\ \citenamefont
  {Bechstedt}(2012)}]{rodl_optical_2012}%
  \BibitemOpen
  \bibfield  {author} {\bibinfo {author} {\bibfnamefont {C.}~\bibnamefont
  {R\"odl}}\ and\ \bibinfo {author} {\bibfnamefont {F.}~\bibnamefont
  {Bechstedt}},\ }\href {\doibase 10.1103/PhysRevB.86.235122} {\bibfield
  {journal} {\bibinfo  {journal} {Phys. Rev. B}\ }\textbf {\bibinfo {volume}
  {86}},\ \bibinfo {pages} {235122} (\bibinfo {year} {2012})}\BibitemShut
  {NoStop}%
\bibitem [{\citenamefont {Onida}\ \emph {et~al.}(2002)\citenamefont {Onida},
  \citenamefont {Reining},\ and\ \citenamefont {Rubio}}]{RevModPhys.74.601}%
  \BibitemOpen
  \bibfield  {author} {\bibinfo {author} {\bibfnamefont {G.}~\bibnamefont
  {Onida}}, \bibinfo {author} {\bibfnamefont {L.}~\bibnamefont {Reining}}, \
  and\ \bibinfo {author} {\bibfnamefont {A.}~\bibnamefont {Rubio}},\ }\href
  {\doibase 10.1103/RevModPhys.74.601} {\bibfield  {journal} {\bibinfo
  {journal} {Rev. Mod. Phys.}\ }\textbf {\bibinfo {volume} {74}},\ \bibinfo
  {pages} {601} (\bibinfo {year} {2002})}\BibitemShut {NoStop}%
\bibitem [{\citenamefont {Refaely-Abramson}\ \emph {et~al.}(2015)\citenamefont
  {Refaely-Abramson}, \citenamefont {Jain}, \citenamefont {Sharifzadeh},
  \citenamefont {Neaton},\ and\ \citenamefont {Kronik}}]{PhysRevB.92.081204}%
  \BibitemOpen
  \bibfield  {author} {\bibinfo {author} {\bibfnamefont {S.}~\bibnamefont
  {Refaely-Abramson}}, \bibinfo {author} {\bibfnamefont {M.}~\bibnamefont
  {Jain}}, \bibinfo {author} {\bibfnamefont {S.}~\bibnamefont {Sharifzadeh}},
  \bibinfo {author} {\bibfnamefont {J.~B.}\ \bibnamefont {Neaton}}, \ and\
  \bibinfo {author} {\bibfnamefont {L.}~\bibnamefont {Kronik}},\ }\href
  {\doibase 10.1103/PhysRevB.92.081204} {\bibfield  {journal} {\bibinfo
  {journal} {Phys. Rev. B}\ }\textbf {\bibinfo {volume} {92}},\ \bibinfo
  {pages} {081204} (\bibinfo {year} {2015})}\BibitemShut {NoStop}%
\bibitem [{\citenamefont {Shih}\ \emph {et~al.}(2012)\citenamefont {Shih},
  \citenamefont {Zhang}, \citenamefont {Zhang},\ and\ \citenamefont
  {Zhang}}]{PhysRevB.85.045132}%
  \BibitemOpen
  \bibfield  {author} {\bibinfo {author} {\bibfnamefont {B.-C.}\ \bibnamefont
  {Shih}}, \bibinfo {author} {\bibfnamefont {Y.}~\bibnamefont {Zhang}},
  \bibinfo {author} {\bibfnamefont {W.}~\bibnamefont {Zhang}}, \ and\ \bibinfo
  {author} {\bibfnamefont {P.}~\bibnamefont {Zhang}},\ }\href {\doibase
  10.1103/PhysRevB.85.045132} {\bibfield  {journal} {\bibinfo  {journal} {Phys.
  Rev. B}\ }\textbf {\bibinfo {volume} {85}},\ \bibinfo {pages} {045132}
  (\bibinfo {year} {2012})}\BibitemShut {NoStop}%
\bibitem [{\citenamefont {Bultmark}\ \emph {et~al.}(2009)\citenamefont
  {Bultmark}, \citenamefont {Cricchio}, \citenamefont {Gr\aa{}n\"as},\ and\
  \citenamefont {Nordstr\"om}}]{PhysRevB.80.035121}%
  \BibitemOpen
  \bibfield  {author} {\bibinfo {author} {\bibfnamefont {F.}~\bibnamefont
  {Bultmark}}, \bibinfo {author} {\bibfnamefont {F.}~\bibnamefont {Cricchio}},
  \bibinfo {author} {\bibfnamefont {O.}~\bibnamefont {Gr\aa{}n\"as}}, \ and\
  \bibinfo {author} {\bibfnamefont {L.}~\bibnamefont {Nordstr\"om}},\ }\href
  {\doibase 10.1103/PhysRevB.80.035121} {\bibfield  {journal} {\bibinfo
  {journal} {Phys. Rev. B}\ }\textbf {\bibinfo {volume} {80}},\ \bibinfo
  {pages} {035121} (\bibinfo {year} {2009})}\BibitemShut {NoStop}%
\end{thebibliography}%
\end{document}